\documentclass[structabstract]{aa} 
\usepackage{graphicx}
\usepackage{txfonts}
\usepackage{tabularx}
\usepackage[squaren,cdot]{SIunits}

\begin{document}
\title{Tests with a Carlina-type diluted telescope}
\subtitle{Primary coherencing}

\author{
  H. Le Coroller \inst{1} \and
  J. Dejonghe    \inst{1} \fnmsep \thanks{First and Second authors contributed equally} \and
  X. Regal       \inst{1} \and
  R. Sottile     \inst{1} \and
  D. Mourard     \inst{2} \and
  D. Ricci       \inst{3} \fnmsep \thanks{FRIA fellow} \and
  O. Lardiere    \inst{4} \and
  A. Le Vansuu   \inst{1} \and
  M. Boer        \inst{1} \and
  M. De Becker   \inst{3} \and
  J-M. Clausse   \inst{2} \and
  C. Guillaume   \inst{1} \and
  J.P. Meunier   \inst{1}}

\institute{ 
  Observatoire de Haute-Provence, F-04870 Saint Michel l'Observatoire, France \\
  \email{herve.lecoroller@oamp.fr; julien.dejonghe@oamp.fr}
  \and 
  Universit\'e Nice-Sophia Antipolis, Observatoire de la C\^{o}te d'Azur,
  CNRS UMR 6525, BP 4229, F-06304 Nice Cedex, France
  \and 
  D\'epartement d'Astrophysique, G\'eophysique et Oc\'eanographie, B\^at. B5C, Sart Tilman,
  Universit\'e de Li\`ege, B-4000 Li\`ege 1, Belgium
  \and 
  AO Laboratory, Mechanical Engineering Department, University of Victoria,
  PO Box 3055 STN CSC, Victoria, BC, V8W 3P6, Canada}

\date{Received Juin 24, 2011}


\abstract
{}
{Studies are under way to propose a new generation of
  post-VLTI interferometers. The Carlina concept studied at the Haute-Provence Observatory is one of the proposed solutions. It consists in an optical interferometer configured like a diluted version of the Arecibo radio telescope: above the diluted primary mirror made of fixed cospherical segments, a helium balloon (or cables suspended between two mountains), carries a gondola containing the focal optics. Since 2003, we have been building a technical demonstrator of this diluted telescope.  First fringes were obtained in May 2004 with two closely-spaced primary segments and a CCD on the focal gondola. We have been testing the whole optical train with three primary mirrors. The main aim of this article is to describe the metrology that we have conceived, and tested under the helium balloon to align the primary mirrors separate by 5-10 m on the ground with an accuracy of a few microns.}
{Getting stellar fringes using delay lines is the main difficulty for
  astronomical interferometers. Carlina does not use delay lines, but the
  primary segments have to be positioned on a sphere i.e. coherencing
  the primary mirrors. As described in this paper, we used a supercontinuum laser source to coherence the primary segments. We characterize the Carlina's performances by testing its whole optical train: servo loop, metrology, and the focal gondola.}
{The servo loop stabilizes the mirror of metrology under the helium balloon with an accuracy better than $5\, \milli\meter$  while it moves horizontally by 30 cm in open loop by 10-20 km/h of wind. We have obtained the white fringes of metrology; i.e., the three mirrors are aligned (cospherized) with an accuracy of $\approx1 \micro\metre$. We show data proving the stability of fringes over 15 minutes,
therefore providing evidence that the mechanical parts are stabilized
within a few microns.  This is an important step
  that demonstrates the feasibility of building a diluted telescope using cables strained between cliffs or under a balloon. Carlina, like the MMT or LBT, could be one of the first members of a new class of telescopes named diluted telescopes.}
{}


\maketitle

\section{Introduction}

Studies are under way to propose a new generation of post-VLTI interferometers.
For example, the OHANA team has proposed to recombine distant
telescopes with optical fibers (Perrin et al. \cite{Perrin1}, Woillez
et al. \cite{Woillez}). Other teams work on new types of combiners
that could provide direct snapshot images, and increase the
sensitivity of numerous aperture interferometers (Tallon \cite{Tallon}, Labeyrie
\cite{Labeyriehyper}, Lardi\`ere et al. \cite{Lardiere1}, Patru et al. \cite{Patru2}). The post-VLTI area will be driven by new fields of research in
astrophysics, such as stellar surface imaging, studies of gravitational
microlensing and central torus of AGN, and exoplanets. To achieve these goals, they
will have to meet several criteria (Ridgway \& Glindemann
\cite{Ridgway}): high angular resolution (Baseline $>100\,\meter$), a good coverage of
uv spatial frequencies (large number of mirrors), and better sensitivity than regular
interferometers (VLTI, Keck, CHARA, etc.). They will be able to accomodate
various focal instruments, such as spectrographs or coronagraphs. A diluted telescope like Carlina could meet all
these criteria. \\

\noindent The optomechanical design of Carlina was described in Le
Coroller et al. (\cite{Lecoroller}; hereafter Paper I). It is similar to the Arecibo
radio telescope, but its spherical primary mirror is diluted (Fig. 3 in Paper I) and
operates at visible wavelengths. There are no delay lines, and it can work with
hundreds of mirrors and should be very efficient thanks to the few intermediate
optical surfaces between the primary mirrors and the final
detector. The stability of the mirrors anchored in the bedrock and an
internal metrology system described in this paper also offer strong
advantages. \\
\noindent The star light is focalized on the focal gondola (at $R/2$) that contains the Mertz (\cite{Mertz})
sphericity corrector, a tracking system, a pupil densifier (Tallon \cite{Tallon}, Labeyrie \cite{Labeyriehyper}),
and a focal instrument such as a photon counting camera (Blazit et al. \cite{Blazit}).  We will describe this complex
focal optical and mechanical set up, its implementation, and the first
observations in a forthcoming paper. In this article, we deal with the metrology at the curvature center of the diluted primary mirror to align (coherencing) the primary mirrors.\\

\noindent In Sect. \ref{goal}, we present the goal of the OHP prototype and the specifications for such a project. In Sect.~\ref{architecture}, we briefly recall the
general principles of Carlina. In Sect.~\ref{OHPProto}, we describe the optical and mechanical solutions that we conceived to build a prototype that responds to the required specifications. The alignment procedure is provided in Sect. \ref{align}. Results are presented at the Sect. \ref{fringes}. We conclude in Sect.~\ref{conclusion} and
propose the idea of a $100\, \meter$ aperture scientific demonstrator named the
large diluted telescope (LDT). A diluted telescope working with hundreds of mirrors
could complement ELTs (D'Odorico \cite{Dodorico}) and
kilometer interferometers (Meisenheimer \cite{Meisenheimer}).


\section{Experiment goals and specifications} \label{goal}

\begin{figure}[h]
  \centering
  \includegraphics[width=8.8cm]{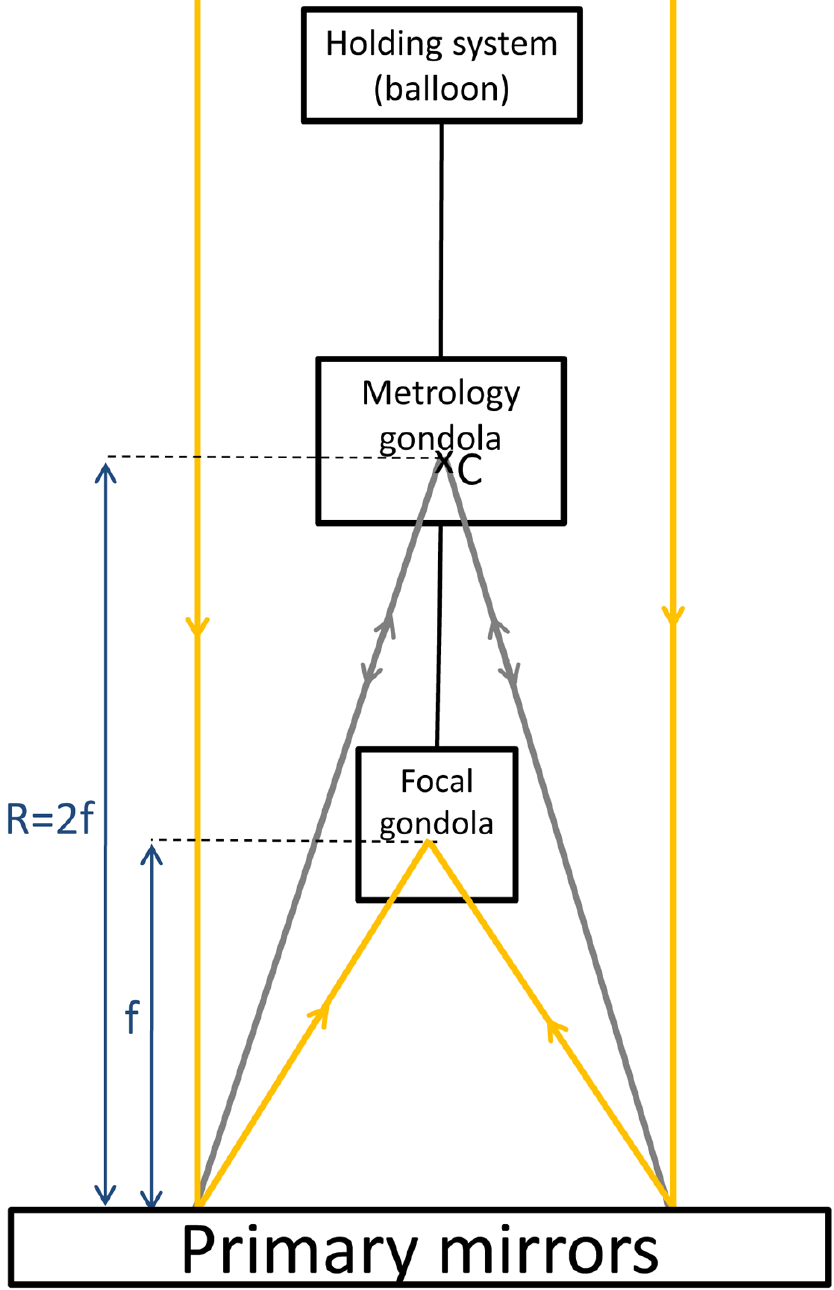}
  \caption{Block diagram of the Carlina project. It is divided in 4 blocks: the primary mirrors, the focal gondola, the metrology gondola, and the holding system that carry the gondolas. The yellow lines show the stellar light, and the gray lines the coherencing laser beam.}
  \label{fonctionnel}
\end{figure}

\noindent The general architecture of Carlina has been detailed in Paper I.
This diluted telescope can be divided in four parts linked with each other as described in the block diagram at Fig. \ref{fonctionnel} and in Sect. \ref{architecture}. These four parts are:\\

\begin{itemize}
 \item A spherical diluted primary mirror made of numerous small mirrors.
\item  A focal gondola that tracks the stars by moving on the focal sphere at the half radius of the spherical diluted primary mirror.
 \item  A metrology gondola at the curvature center of the diluted primary mirror. It is used to adjust the mirrors position on the primary sphere with high accuracy (primary coherencing). A part of this metrology gondola (source, and CCD) can be placed on the ground (Sect. \ref{metrology}).
 \item  A holding system that carries the metrology and focal gondola.
\end{itemize}

   \noindent The main goal of the experiment described in this paper is to obtain stable metrology fringes in order to align the primary mirrors with one micron accuracy i.e. we have to obtain fringes that stay in the field of view of the metrology CCD and that move  slowly enough to be frozen in a short time exposure ($\approx1$ millisecond).
   We provide the specifications in order to reach this goal in Appendix \ref{CarlinaSpec}. In the next sections, we describe in detail the Carlina architecture and the solutions found with the OHP prototype in order to answer to the specifications of Table \ref{tablspec}.  We will see that the OHP prototype reaches perfectly  these specifications (last column of Table \ref{tablspec}). The focal gondola characteristics will however be described in a forthcoming paper.

\begin{figure*}[t]
  \centering
  \includegraphics[width=15cm]{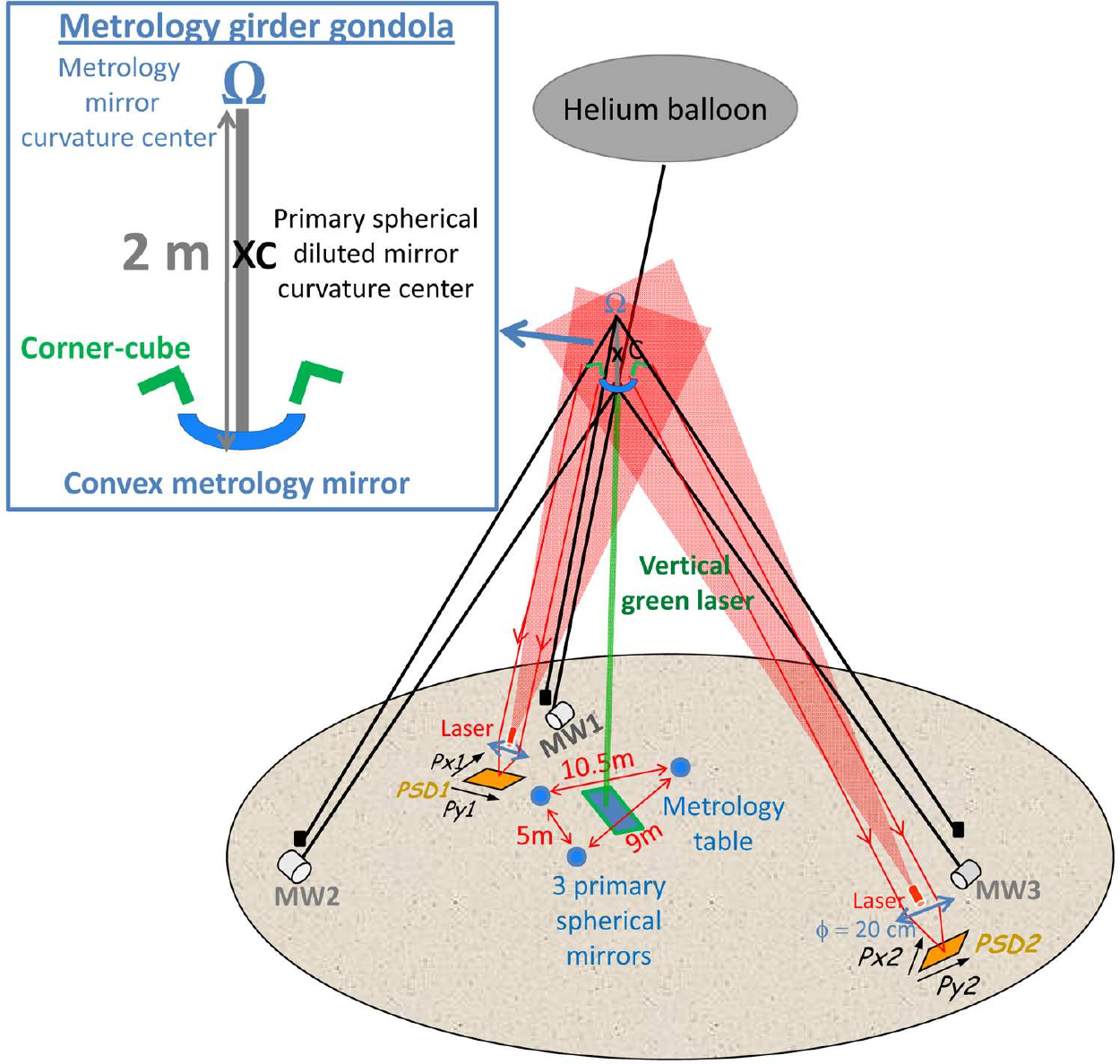}
  \caption{Schema of the OHP interferometer prototype. In order to show the whole
    experiment, this is not a scale drawing. The black lines represent the cables. The blue circles
    symbolize the three primary mirrors at the ground level around the
    metrology table (Fig.~\ref{photoMirroirTable}). The convex
    metrology mirror is attached at the bottom of a girder
    gondola. The helium balloon keeps the cables under tension by
    pulling to the lower part of this girder gondola. Point $C$ which corresponds to
    the center of the spherical diluted primary mirror, is at the
    middle of the girder. $\Omega$ at the top of the girder is the
    curvature center of the convex metrology mirror. It is passively
    positioned by three cables attached to the ground. This mechanical
    design allows the metrology mirror to rotate around its own center
    of curvature $\Omega$, if the balloon oscillates under the wind. The
    white cylinders ($MW1$, $MW2$, $MW3$) are the three motorized
    winches stabilizing the metrology mirror. The devices to measure the displacements of the metrology gondola are also
    shown: red lasers in front of the $20\, \centi\meter$ lens and
    corner cubes attached on the metrology gondola
    (Sect.~\ref{servoloop}). For the sake of clarity, the focal gondola under the metrology gondola is not represented.}
  \label{shematCarlinaProto}
\end{figure*}

\section{General principles}\label{architecture}

 A cable tripod tensioned by a helium balloon, or cables attached between two mountains ensures that the suspended optics is stabilized  within the limits defined by the cable sag. The top of this tripod is also the curvature center of the
spherical diluted primary mirror (Fig. 3 in Paper I). This point is used for the alignment of the primary segments using a metrology gondola. 
\noindent Under the metrology gondola, a focal
gondola is constrained by cables to move along the half radius focal
sphere.
\noindent The main advantages of such a design are the absence of delay lines,
and the simplicity of the optical train.

\section{Specificities of the OHP prototype and description of its subsystems}
\label{OHPProto}

\noindent One goal of the $10\, \meter$ baseline OHP prototype is
to test the whole optical train of Carlina. In the next sections, we describe the optical and mechanical solutions that we found to build a prototype (holding system, metrology gondola, and primary mirror). We discuss the technical choices  in order to respond to the required specifications of Table \ref{tablspec}. All the subsystems described in this article are schematically represented in Fig. \ref{shematCarlinaProto} that gives an overview of the complete experiment (without the focal gondola).

\subsection{Optical design}

\subsubsection{Baselines and position of the mirrors}

The CARLINA design at OHP has a $R=71.2 \, \meter$ curvature radius ($R$ is constrained by the maximum space available at OHP for installing the holding system). The focal gondola is on the focal sphere at $\frac{R}{2} = 35.6 \,\meter$. The classical corrector of spherical aberration has a maximum aperture ratio equal to $f/2$ (Paper I). It means that the maximum baseline at Haute-Provence observatory is $18\,\meter$. We decided to start with three mirrors on the ground, forming three baselines of
respectively $5$, $9$, and $10.5\, \meter$ (Fig.~\ref{shematCarlinaProto}). The 9 m baseline is oriented N-S and the smaller baseline E-W. These baselines constitute a reasonable intermediate step towards the mirrors of a future 100 m aperture scientific project.

\subsubsection{Description of the metrology system}\label{metrology}

\noindent The primary mirrors
have to be aligned (on a sphere of curvature radius $R=71.2 \, \meter$ for the OHP prototype) with an accuracy about equal to the atmospheric
piston, typically one micron. This is the goal of the metrology
described in this section.\\

\noindent The principle: the primary segments
produce an image of a point source located at their common center of
curvature in the metrology gondola (Fig. \ref{fonctionnel}); we adjust the tilt-tip of each mirror by superimposing
these images at the curvature center; the piston is adjusted by searching
for fringes. The residual piston errors are comparable to the
coherence length of the light source and it is adjusted down to the
seeing limited value of a few microns, using a source of white light.\\

\begin{figure}[!t]
  \centering
  \includegraphics[width=8.8cm]{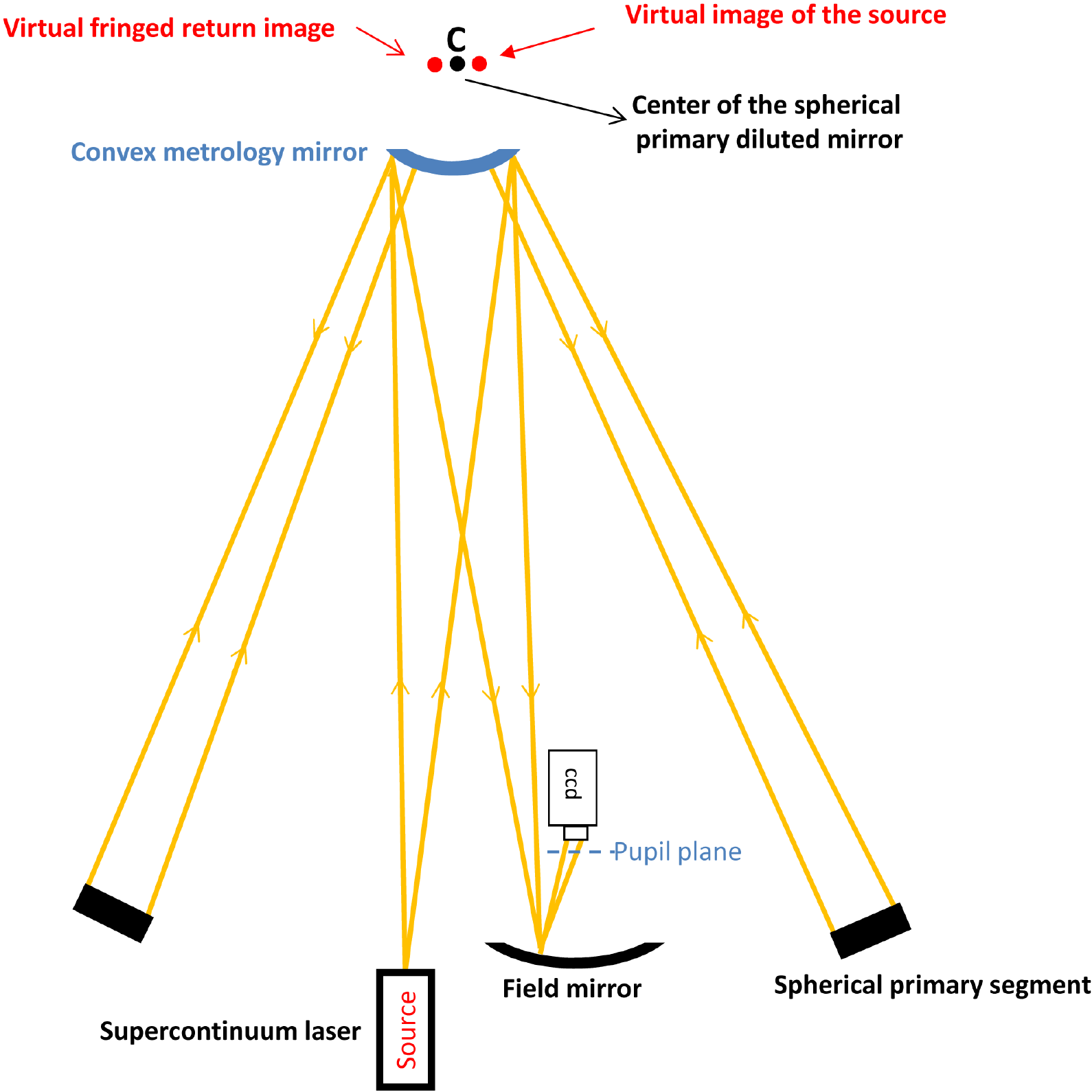}
  \caption{Schema describing the coherencing technique. A laser source
    lights up a convex mirror near the curvature center of the
    spherical diluted primary mirror. The convex metrology mirror
   creates a virtual image of this source that is seen as a
    point source by the primary segments. The light comes back on a
    field mirror where fringes are formed. A CCD is placed a few
    centimeters after the exit pupil plane, where the diffraction of
    the small subapertures  is high enough to observe the fringes (the
    beams overlap). This optical design tolerates translation ($\pm
    2\, \milli\meter$ with the OHP prototype optics) of the metrology mirror at the curvature
    center. With a broadband laser source, finding the white fringe
    allows us to adjust the piston balance within one micron.
    On the prototype at OHP, the laser source, the field mirror and the CCD are placed on a metrology table (Fig. \ref{tableOptiquephoto}).}
  \label{shematMetrology}
\end{figure}

\noindent In practice, the light source and the camera at the
curvature center have been replaced by a
convex mirror called a metrology mirror (Fig. ~\ref{shematMetrology}).
The convex metrology mirror creates a virtual image (near
point $C$ on Figs. \ref{shematCarlinaProto}, \ref{shematMetrology})
 of a light spot produced by a laser located on
a metrology table at ground level (Sect.~\ref{metrologytable},
Fig.~\ref{tableOptiquephoto}), itself seen as a point source by the
primary segments. The primary segments create a virtual fringed return
image. The light comes back toward a
field mirror on the metrology table where fringes are formed.
 A CCD is placed a few centimeters after
the exit pupil created by the field mirror, where beams are partially
superimposed due to diffraction. This optical solution allows lighter equipment to be carried under the helium balloon (only one mirror). It is simpler to use, because there is no need for energy and no need to pilot a camera and lasers from the ground.\\

\subsubsection{Design of the metrology optics}\label{zemax}

\begin{figure*}[!t]
  \centering
  \includegraphics[width=15cm]{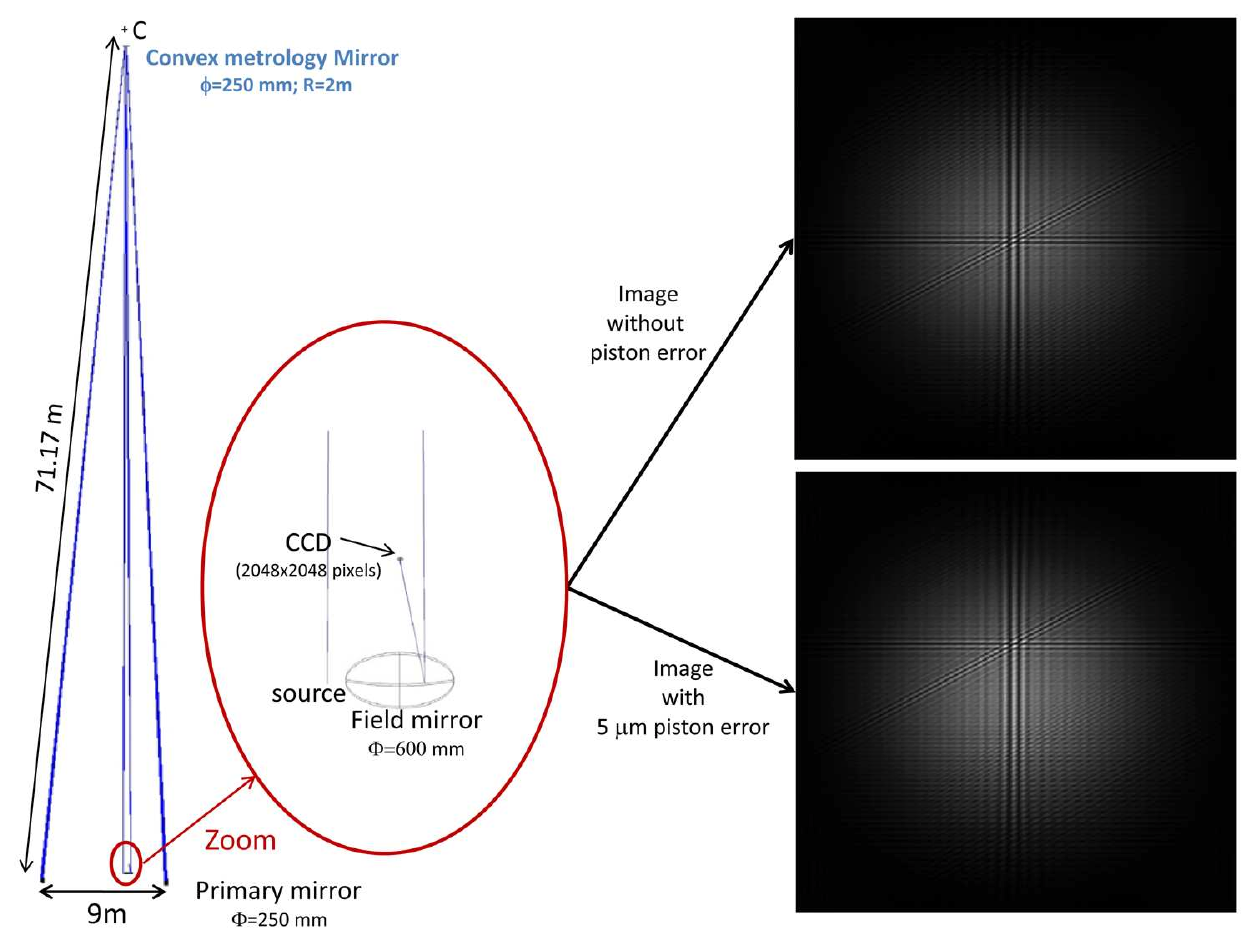}
  \caption{Optical design of the OHP prototype metrology simulated with the
    Zemax software. On the left, a scale schematic view similar to
    that shown in Fig.~\ref{shematMetrology} illustrates the plane
    including the $9\, \meter$ baseline and the $C$ point.  In the
    middle, a zoom of the field mirror is shown: the ray tracing is
    for a $1 \milli\meter$ offcenter metrology mirror (horizontal displacement). On the right,
    the two images give the results of simulations performed,
    respectively, without and with piston errors (on the N mirror). In the two cases, the
    fringes of the three baselines at OHP ($5 \meter$, $9 \meter$, $10.5
    \meter$) are shown.}
  \label{simuzemax}
\end{figure*}

The metrology optics to cospherize the primary segments
(schematically described in Fig.~\ref{shematMetrology}) has been
designed with the Zemax software (Fig.~\ref{simuzemax}) in order to
accept the oscillations of the convex mirror at the curvature center
of the primary ($\pm\, 2 \,\milli\meter$ measured experimentally at the top of a tripod attached to a balloon in a relatively constant wind). We then found that a 600 mm field mirror is required (for metrology mirror of $1\,\meter$ focal length): the fringes cross the 600 mm field mirror (Fig.~\ref{simuzemax}) with horizontal oscillations of $\pm \,2\,\milli\meter$ of the metrology mirror in the wind. We will see that the field mirror has been slightly oversized because we attached the metrology mirror in a such a way that it turns about its own curvature center (Sect. \ref{optomechanical}): the horizontal displacements of  $\pm\, 2\, mm$ are transformed in rotations.

\begin{figure}[h]
  \centering
  \includegraphics[width=5cm]{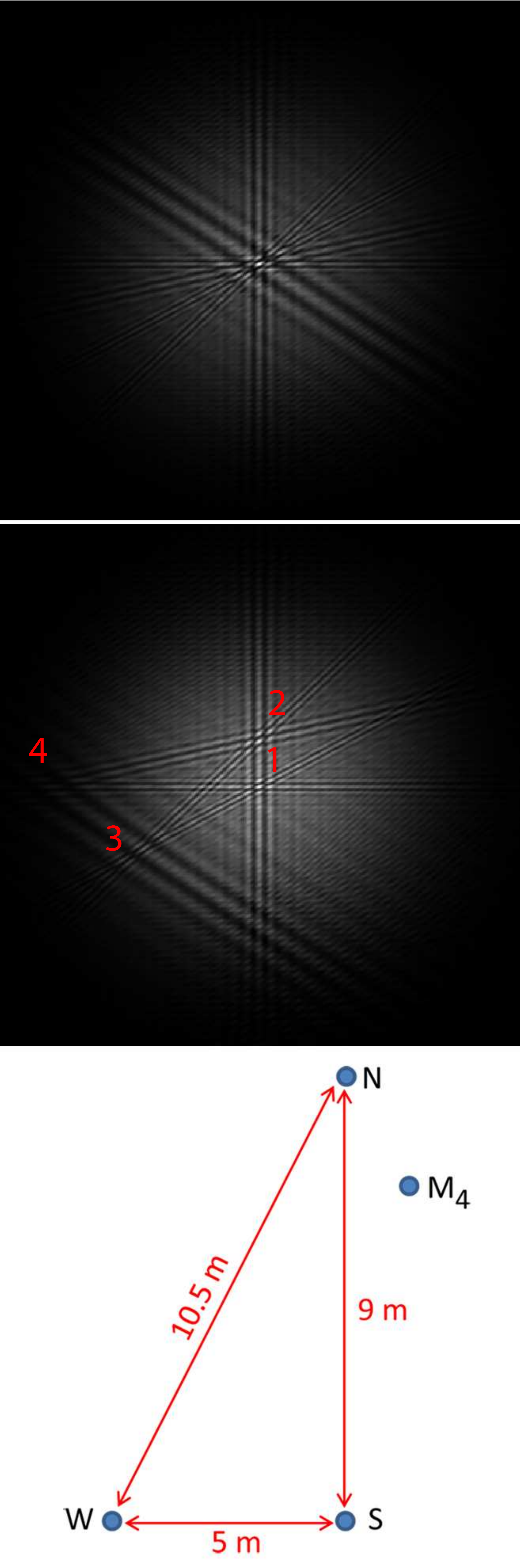}
  \caption{Same simulation as in Fig. \ref{simuzemax} but with a fourth mirror [if n is the mirror numbers: $n(n-1)/2=6$ fringes and $C_4^3=4$ triplets]. Top: all the mirrors are in phase. Middle: the three mirrors of Fig. \ref{simuzemax} are still in phase, but a 3 micron piston error has been added on the fourth mirror $M_4$. We wrote numbers in red, which are close the intersection of each triplet of fringes. The number 1 corresponds to the triplet of fringes without piston error of Fig. \ref{simuzemax}. Bottom: Position at scale of the four mirrors.}
  \label{franges4mirroirs}
\end{figure}

\noindent In a conventional optical design, we would place a lens in
front of the CCD (Figs.~\ref{shematMetrology}, \ref{simuzemax}) to
produce an image of the fringes (field mirror), but in our case there are no optics between the field mirror and the
camera. Fringes appear quasistatic on the camera with this optical design because they are in a ``pseudo pupil plane'' a few
centimeters after the exit pupil where light beams are
partially superimposed, as diffraction gives rise to significant beam
overlap (practically total). For
future projects, a tip-tilt mirror could be placed at the exit pupil
created by the field mirror, to stabilize the image at the entrance of
a ``dispersed speckle'' piston sensor (Borkowski et
al. \cite{Borkowski}; Tarmoul et al. \cite{Tarmoul}) or any other
``wavefront sensor'' adapted to diluted pupils. \\
\noindent Three sets of fringes always intersect each other at a unique common
point (top and bottom images of the Fig. \ref{simuzemax}). 
Along the line defined by a white fringe, the
light beams coming from two apertures are in phase. For mirrors N and S (see position of the OHP mirrors on Fig. \ref{photoMirroirTable}),
this translates into the following identity: $\phi_N = \phi_S$ (where
$\phi$ is the phase of the light beam from a given mirror). The same
principle applies for the white fringe created by the mirrors N and W:
$\phi_N = \phi_W$. At the intersection between of both white fringes,
$\phi_N = \phi_S = \phi_W$. The white fringe created by mirrors S and W
(defined by the line where $\phi_S = \phi_W$) therefore passes  through
this intersection. Fig. \ref{franges4mirroirs} shows the results of the same study as presented in Fig. \ref{simuzemax} but with a fourth primary mirror. The fourth mirror is aligned by adjusting its piston screw, until the three sets of fringes linked with this mirror cross the reference triplet   (centered on the envelope as described at Fig. \ref{simuzemax}) in a unique point. This technique is very sensitive because the fringes are very narrow. Without turbulence (for example if we build such an interferometer in space), it should be possible to adjust their position with respect to each other with  much better accuracy than the width of one fringe (piston error $<<1\,\micro\meter$).\\

\begin{figure}[h]
  \centering
  \includegraphics[width=5 cm]{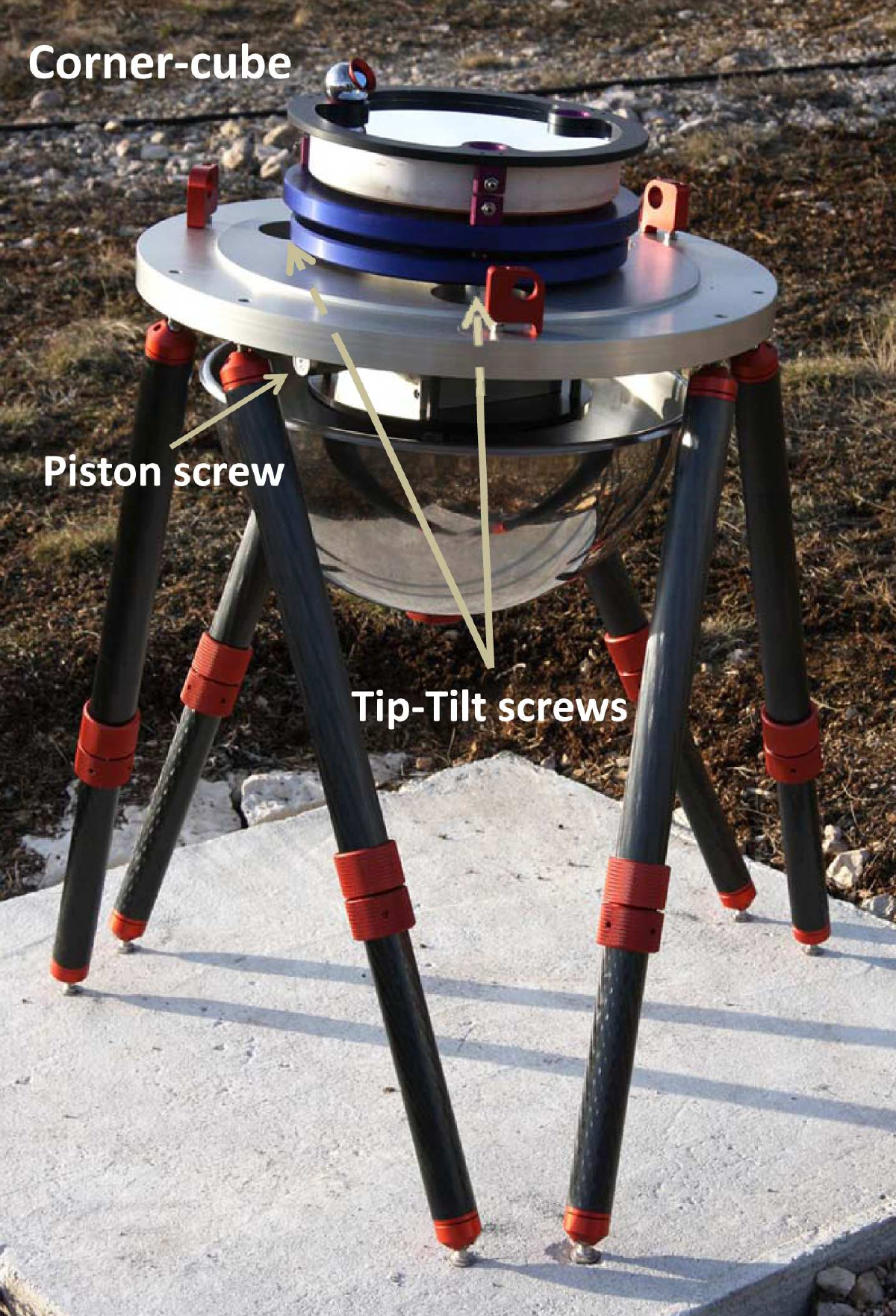}
  \caption{Picture of the mount that carries one of the primary mirrors.  With
    this hexapod we can position the mirror in rotation and
    translation.  The carbon fiber legs insure a stability better
    than one micron. }
  \label{hexapod}
\end{figure}

\noindent Usually, regular interferometers are densified. The
densification shrinks the diffraction envelope. Consequently, the
white fringe moves out of the diffraction envelope for a piston error
of only a few microns. Fringes are spectrally dispersed (Weigelt et
al. \cite{Weigelt}, Petrov et al. \cite{Petrov}, Mourard et
al. \cite{Mourard}, Tarmoul et al. \cite{Tarmoul}) both for scientific
reasons, and also in order to increase the coherence length in order to be able to detect them even with a large piston error.
 In our case, we did not need to spectrally disperse the
fringes because our metrology works in ``Fizeau'' mode, and the white
fringe can be detected in a large speckle envelope without spectral
dispersion, i.e. $\approx 50\, \micro\meter$ piston error on the
primary mirrors at OHP. This is an important and interesting solution of our metrology
optical design to coherence the mirrors of an interferometer (see also discussion on the CCFFS in Sect. \ref{fringes}).


\subsubsection{Optical path errors that are not common to focal beam and to metrology beam}
\label{noncommon}

We have described a metrology at the curvature center. In this section, we explain how it is ensured that, if the primary mirrors are coherenced from the curvature center, they are also coherenced for the stellar light. Typically, the accuracy
of the metrology at the curvature center (to align the primary mirrors),
is equal to the wavelength used for the metrology. The metrology indeed
proceeds through the measurement of fringes for which a
displacement of one fringe corresponds to an error of one lambda on the
wavefront. Also, to be sure that the measured coherencing level at the curvature center is sufficient for observations of stellar objects, it is preferable to use a metrology source with an effective wavelength smaller than or equal to the effective wavelength of the stellar light (middle of the filter bandwidth). By using the same wavelengths for the metrology and the stellar light, an error of one lambda on the wavefront (piston error of $\lambda/2$ for a primary mirror) moves the fringes pattern by one metrology fringe at the curvature center and one stellar fringe at the focal camera. Nevertheless, if we observe in the visible we gain nothing by using a metrology source with an effective wavelength that is much smaller than one micron because the atmospheric turbulence will add more than one micron of piston error to the stellar light.  \\

\noindent Also there is no optical surfaces between the primary and the focal gondola that could introduce optical path errors, which would not be seen by the metrology at the curvature center. Only the atmosphere adds a piston error that is not measured by the metrology at the curvature center because the star light does not cross the same atmospheric layers (Fig. \ref{fonctionnel}). In the focal gondola, the corrector of spherical aberration (Mertz \cite{Mertz}) is made of two aspheric mirrors.  This device is made of continuous optical surfaces (no independent optics for each beam), and supposes a spherical primary, which is a condition ensured by the metrology at the curvature center. A diamond turning machine has shaped these mirrors with an accuracy much better than one micron, i.e., better than the atmospheric piston error. \\ 
To conclude, it is ensured that if the mirrors are ``coherenced" (cospherized) from the curvature center, they are also ``coherenced" for the stellar light with about the same accuracy (the atmospheric piston $\approx \, 1 \micro\meter$).  This metrology is equivalent to the active optics of a regular telescope. From this point of view, Carlina looks like a regular telescope working without delay lines. When the primary segments of a telescope are aligned on a parabolic surface or the surface adapted to the conjugation of the telescope mirrors (in our case, a spherical primary + Mertz corrector in the focal gondola) with an accuracy of a few $\lambda$, the stellar light is coherenced at the focal point.


\subsection{Mechanical design} \label{mecanic}

\subsubsection{The mount of the primaries mirrors}

 Each mirror is supported by an hexapod mount made of carbon fiber legs that ensure micrometric stability (Fig. \ref{hexapod}). An hexapod has also six freedom degrees that allows adjusting the position of the mirror in rotation and translation. The length of each leg can be adjusted and blocked to position the mirror with one millimeter accuracy. A second micrometric stage allows finely adjusting the position of each mirror (Fig. \ref{hexapod}): two micrometric screws control the
tip-tilt, and the piston is adjusted using a vertical translation
stage under each mirror.  Thus, nine parameters
(screws) have to be adjusted within one micron to cospherize the
three mirrors around a common curvature center ($C$ on
Figs.~\ref{shematCarlinaProto}, \ref{shematMetrology}, \ref{simuzemax}).

\subsubsection{Holding system}

\noindent At OHP, the focal gondola is at $f=35.6\, \meter$ and the metrology gondola at $R=71.2\, \meter$ above the ground. A 80 to 100 m high pylon costs about 100 Keuros, that is too expensive for a demonstrator. We then decided to use a captive helium balloon to attach the gondolas above the OHP ground.   We are in the most difficult conditions due to the wind resistance of the balloon. The balloon is also a possible solution for a future very large diluted telescope of 500 m or more. In order to minimize the oscillations of the balloon, it must have a low resistance in the wind with the maximum traction. We used a 12 m long balloon with eliptic shape (the maximum size for the room where the balloon is stocked). This balloon has a $70\,\kilogram$ payload.\\
\noindent The gondolas are suspended under the helium balloon with $1.8\,
\milli\meter$ diameter cables, made of ``PBO'' fibers (Orndoff \cite{Orndoff}), a member of
the polybenzoxazole class of polymers. They have higher tensile
strength and Young's modulus than Kevlar: the PBO HM Young's modulus is about $3.8$ times that of the Kevlar29 Young's
modulus corresponding to $270\, \giga\pascal$ for about the same density ($1.5\,
\gram/\centi\meter^3$). The total cables weight is about 14 $\%$ of the balloon payload.


\subsubsection{The metrology gondola}
\label{optomechanical}

Experimentally, we know that the top of the tripod $71.2\,\meter$ above the ground can move of several centimeters when wind direction changes, and oscillate horizontally of $\pm 2\, \milli\meter$ in the wind (using $1.8\,
\milli\meter$ PBO cables).
If the metrology mirror has a horizontal translation movement exceeding $\pm2 \, \milli\meter$ the light goes out of the field mirror (Sect. \ref{zemax}). We then studied a mechanical way to attach the metrology
mirror  to limit the oscillation effects (due to the balloon movements in the wind) on the optical
beams (Fig.~\ref{shematCarlinaProto}). As this one is nearly spherical, if the metrology mirror rotates around its curvature center, the return beams doesn't move on the field mirror on the ground. The design of the gondola allows the translation motions of the metrology mirror (if the balloon oscillates) to be converted into  pure rotation around its own curvature center, because it is attached under a girder gondola
 along its radius of curvature. The length ($2\meter$) of this girder is equal to the
curvature radius of the metrology mirror (Fig.~\ref{photoGondoGirder}).
Two cable tripods are attached at the bottom
and at the top of this metrology gondola (Figs.~\ref{shematCarlinaProto} and ~\ref{photoGondoGirder}).

\begin{figure*}[!t]
  \centering
  \includegraphics[width=15cm]{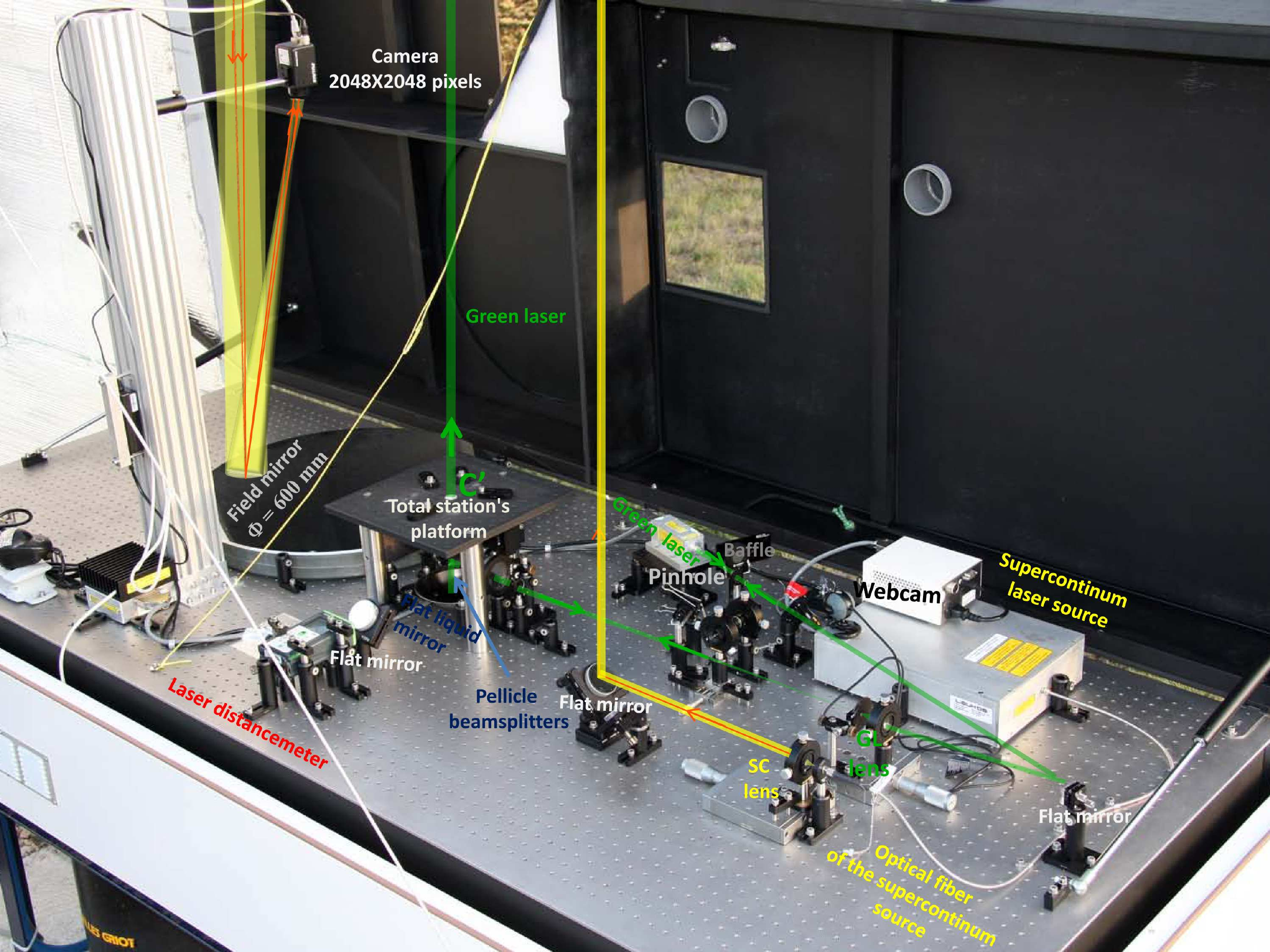}
  \caption{Picture of the metrology table. A large white hood (on the
    left in the picture) protects the table from the rain.  To protect
    the optics from humidity and wind, a second black hood is also used
    (open on this photo to see the optics) when we are making
    observations.  It has several holes to allow laser beams to pass
    through. A green laser is used as a vertical reference (a principle schema is presented Fig. \ref{laservertical}) in order to align the metrology gondola. The laser distance meter measures the altitude of the
    metrology gondola to a precision of $\approx 1 \milli\meter$.  As
    described in Fig.~\ref{shematMetrology} and Fig.~\ref{simuzemax}, we use the supercontinuum
    laser to cospherize the primary mirrors; a camera above the
    $600\, \milli\meter$ field mirror near a pupil plane records the
    white fringed return image (only visible when the primary segments are
    aligned).}
  \label{tableOptiquephoto}
\end{figure*}

\noindent The top of the girder is maintained by an unmotorized tripod (yellow upper tripod in Fig. \ref{photoGondoGirder}). The bottom of the girder carries
the metrology mirror and is attached to the balloon pulling on a
motorized lower tripod (Figs.~\ref{shematCarlinaProto}
and~\ref{photoGondoGirder}).  The cables that hold the balloon go
through a spacing triangle preventing them from touching the
girder gondola.  As the varying forces due to balloon
oscillations are applied to the girder's bottom, its top end $\Omega$
remains quasistatic, and the metrology mirror rotates around its
curvature center $\Omega$, with a reduced effect on light beams. Nevertheless, if the girder pushes to the side, $\Omega$ moves slightly because the forces equilibrium in the top tripod is modified. \\
\noindent In passive mode (when the lower tripod motors are not running), the residual
motions of the girder's bottom are $5$--$30\, \centi\meter$, and less
than $1\, \centi\meter$ at the top $\Omega$. Obviously, these residual movements produce negative
effects:
\begin{itemize}
\item The metrology mirror can move
outside the beam if the balloon pulls strongly sideways.

\item The metrology fringes fall outside the field mirror
  (Fig.~\ref{simuzemax}) if the metrology mirror curvature
  center $\Omega$ is more than $2\, \milli\meter$ offcentered.
\item High velocity vibrations can
  scramble the metrology fringe detection.
\item The tripod's motion induces oscillations of the focal gondola
  that perturb the guiding and scramble the observed stellar fringes.
\end{itemize}

\noindent To solve these problems, the metrology mirror (girder's
bottom) is actively stabilized (sect. \ref{servoloop}).

\begin{figure*}[!t]
  \centering
  \includegraphics[width=15cm]{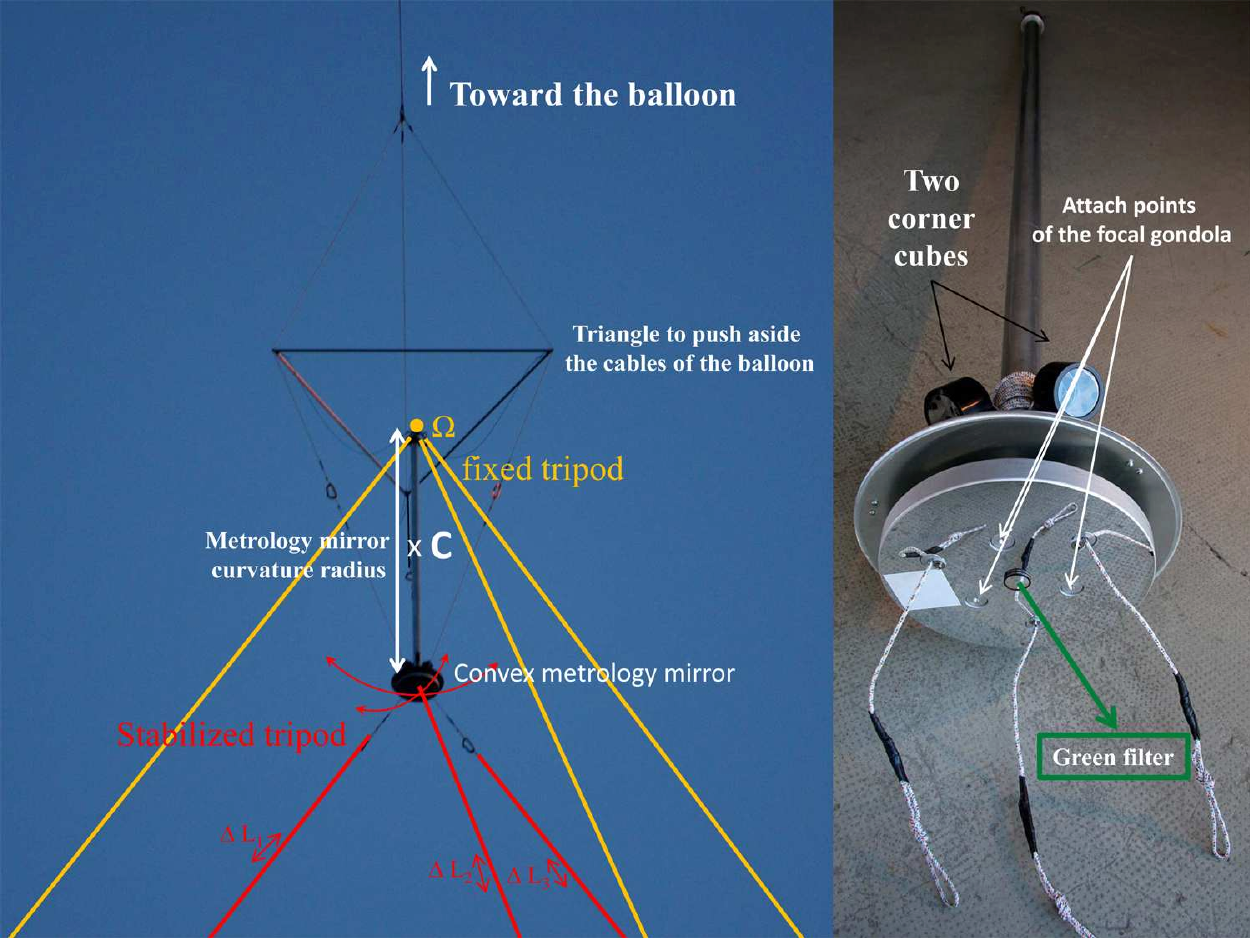}
  \caption{On the left is shown a photograph of the girder metrology gondola at the
    curvature center of the diluted primary mirror.  We drew in yellow
    and red the cables with low contrast on the picture. The convex
    metrology mirror is attached in such a way that it mainly turns
    around his own curvature center $\Omega$ when the balloon
    oscillates in the wind.  The bottom tripod (in red) is stabilized.
    On the right, a close view of the girder gondola: two corner cubes
    attached at the bottom of the girder are used by the servo loop
    system (Sect. \ref{servoloop}). The green filter at the center of
    the mirror creates a shadow on the metrology table on the
    ground. It improves the contrast of the white fringe return
    image. Behind the green filter, a circular sight is used to align
    vertically the gondola.  Six holes have been drilled in the
    metrology mirror using a diamond tool.  The bottom tripod cables (red)
    converge through three holes toward the point where the balloon
    is attached.  The three attachment points of the focal gondola are
    oriented toward the middle of the girder ($C$ point) in such a way
    that it will turn around the curvature center of the diluted
    primary mirror.  The laser distance meter's light (Fig. \ref{tableOptiquephoto}) is reflected on
    the white sheet on the left of the metrology mirror.}
  \label{photoGondoGirder}
\end{figure*}

\subsubsection{Metrology table}
\label{metrologytable}

The experiment (girder metrology gondola and primary mirrors) has been aligned from a stable metrology table (Figs. \ref{tableOptiquephoto}, \ref{photoMirroirTable}). This table is equipped with:\\

\begin{itemize}
  \item A standard high precision industrial measurement device (ZEISS total station) to position the mirrors. The so called total station consists in a theodolite combined with a laser distance meter.
\item A vertical green laser used as a reference to align the metrology gondola.
 \item A LEUKOS-SM (class 3b) supercontinuum laser source (for alignment of the primary mirrors with one micron accuracy) delivering 96 mW (pulse width $<1$ ns) with a wide spectrum from $300\, \nano\meter$ to $2400\, \nano\meter$. IR has been removed with a wide band hot mirror (reflecting $\approx 100 \%$ of the light beyond $750 \,\nano\meter$) placed at the fiber output.
\item A laser distance meter to measure the altitude of the metrology gondola.
\end{itemize}

\noindent In the next sections, we describe how we have aligned all the optics from the metrology table.

\section{Alignment} \label{align}

\noindent All the optics and mechanical parts have been positioned
from the point ($C'$) defined as the vertical projection of the curvature center of the
primary sphere (Point $C$ on Figs.{~\ref{shematCarlinaProto}, ~\ref{shematMetrology},
and~\ref{simuzemax}) onto the metrology table on
the ground (Fig. \ref{tableOptiquephoto}).

\subsection{Alignment of the metrology gondola}
\label{alignmetrogondol}

\noindent The metrology gondola is positioned vertically above $C'$
thanks to the vertical green laser of the metrology table (Figs.~\ref{tableOptiquephoto}, \ref{laservertical}).
The green laser is vertically adjusted within one to two arcsec
thanks to a silicon flat liquid mirror on the metrology table (Fig.~\ref{tableOptiquephoto}).
A few centimeters above this liquid mirror, a tilted pellicle beam
splitter (inclination at $45\degree$) reflects the light of the
$150\, \milli\watt$ green laser ($\lambda = 532 \nano\meter$) exactly
toward its own source in such a way that the other part of the beam
reflected on the liquid mirror goes vertically to the metrology
gondola (Fig.~\ref{shematCarlinaProto}).\\

\noindent Initially, we
observe the metrology mirror ($\approx 71 \meter$ above the metrology
table) with a small telescope located on the side of the metrology table, and
we adjust the length of the lower tripod cables (red cables in Fig. \ref{photoGondoGirder}) until the middle of
the mirror approximately intersects the green laser beam
(Figs.~\ref{shematCarlinaProto} and~\ref{photoGondoGirder} ). The green
laser is focused from the metrology table (using the GL Lens of
Fig.~\ref{tableOptiquephoto}) to the center of the metrology mirror
($\Omega$ in Fig.~\ref{photoGondoGirder}) so that the return spot
appears as small as possible on the ground. Then, we adjust the cable
length of the top tripod (yellow cables in Fig. \ref{photoGondoGirder}) to send back the green laser light toward
$C'$ to ensure that the metrology mirror curvature center
$\Omega$ is accurately positioned vertically above $C'$. A
sight is drawn on the center of the metrology mirror disk
(Fig.~\ref{photoGondoGirder}), and a diffractive image of this sight
is thus visible in the return spot. An accurate alignment is obtained by manipulating the lower tripod's
motorized winches, to center this image in the return spot (the
sight is vertically positioned above $C'$), ensuring that the girder
gondola is perfectly vertical.
\noindent A laser distance meter measures the altitude of the
metrology mirror with one millimeter accuracy
(Figs.~\ref{tableOptiquephoto} and ~\ref{photoGondoGirder}). The
vertical tolerance is $\Delta_z=5 \, \milli\meter$ for $10\,\meter$
baselines and the atmospheric turbulence at OHP (Table \ref{tablspec}). We iterate several
times this process of alignment until the metrology gondola is
vertically oriented above the metrology table and at the correct
altitude.\\
\noindent Once the metrology gondola is correctly positioned, we turn
on a servo loop.

\subsubsection{Servo loop system to maintain the metrology gondola alignment}\label{servoloop}

With a wireless anemometer attached to the balloon, we measured (as seen from the ground with a small telescope) that an horizontal wind speed up to $15\, km/h$ shifts horizontally the metrology mirror of about $30\, cm$. These large movements are slow, due to balloon inertia ($30\, \second$ to a few minutes). Wind buffeting at typically $1\,\hertz$ induces displacement of a few millimeters. We have developed a Servo loop to correct for these motion. Note that above $\approx 20$ km/h wind speed, the experiment goes down (a cable of the tripod is slackened) and it is not possible to work even with a servo loop. This result could probably be improved a lot using a balloon with better aerodynamical behaviors.\\

\noindent The stability of the metrology gondola is achieved by means
of the lower tripod of the girder (red cables in Fig. \ref{photoGondoGirder}).  The lengths of the tripod cables
are accurately controled by three computer driven winches ($MW1$,
$MW2$, $MW3$ in Fig.~\ref{shematCarlinaProto}). At ground level, two red lasers light up two cornercubes (Fig. \ref{photoGondoGirder}) attached on the girder gondola (close to the metrology mirror). The
cornercubes return the light toward the sources
(Fig.~\ref{shematCarlinaProto}), and the $20\, \centi\meter$ large lens,
behind each laser creates an image of the cornercubes on a position
sensitive detector (PSD).  We measured a $50\, \milli\second$ time response to longitudinal constraints for a 100 m Kevlar cable (Appendix \ref{cable}). We therefore
operate the servo loop at $10\,\hertz$: Each PSD gives two coordinates ($P_x$,
$P_y$) of the image photocenter sampled at $300\,\hertz$ and down
sampled at $10 \,\hertz$ after passing through an antialiasing
filter. An algorithm determines the girder position (position of the
corner cubes). Three computer-controlled winches bring back the girder toward its initial position ($MW1$, $MW2$, $MW3$
in Fig.~\ref{shematCarlinaProto}).\\

\noindent We use an algorithm derived from adaptive optics devices (Jacubowiez
et al. \cite{Jacubowiez}).  It is a method based on a singular value
decomposition (SVD) of the rectangular matrices. Such a method works
even if the positions of the motor winches and of the girder gondola
are not a priori fully determined.

\noindent The lower tripod summit motions (metrology mirror in Fig. \ref{photoGondoGirder})
are linked with the displacements of the photocenters (image of the
corner cube) on both PSDs (Fig.~\ref{shematCarlinaProto}) by the
linear equation system:
\begin{equation}
  \Delta\vec{L}=
  M\cdot \Delta \vec{P}=
  \pmatrix{
    \Delta L_{1}\cr
    \Delta L_{2}\cr
    \Delta L_{3}\cr
  }=
  \pmatrix{
    m_{11}&m_{12}&m_{13}&m_{14}\cr
    m_{21}&m_{22}&m_{23}&m_{24}\cr
    m_{31}&m_{32}&m_{33}&m_{34}\cr
  }
  \pmatrix{
    \Delta P_{x1}\cr
    \Delta P_{y1}\cr
    \Delta P_{x2}\cr
    \Delta P_{y2}\cr
  }
   \nonumber
\end{equation}
\noindent Determination of the interaction matrix $M$ is
described in appendix \ref{systemcalib}. It is used in a velocity loop 
proportional integral derivative (PID) to drive the three
motor winches at $10 \hertz$:
\begin{equation}
  \vec{V}(n)=
  K_p\cdot\Delta\vec{L}(n)+
  K_I\sum_{i=1}^{n} \Delta \vec{L}(i)+K_d[
  \Delta \vec{L}(n)-\Delta \vec{L}(n-1)
  ]  
\end{equation}
where n is the discrete step at time t, $\vec{V}$ the command velocity
sent to the motor winches, $K_p$ the proportional gain, $K_I$ the
integral gain, and $K_d$ the derivative gain.\\

\noindent In active mode, bottom (metrology mirror in Fig. \ref{photoGondoGirder}) and top ($\Omega$ in Fig. \ref{photoGondoGirder}) movements of the girder are reduced respectively
within $\approx 5 \,\milli\meter$ and $\approx 0.3 \,\milli\meter$ (see Sect. \ref{fringes}). The residual servo loop errors are mainly due to the gondola inertia, the cable response (delay for the gondola to move when we pull on a cable), the atmospheric turbulence (which limits the accuracy with which the position of the
 corner cubes can be measured), the electronic noises, etc.\\

\begin{figure}[h]
 \includegraphics[width=7cm]{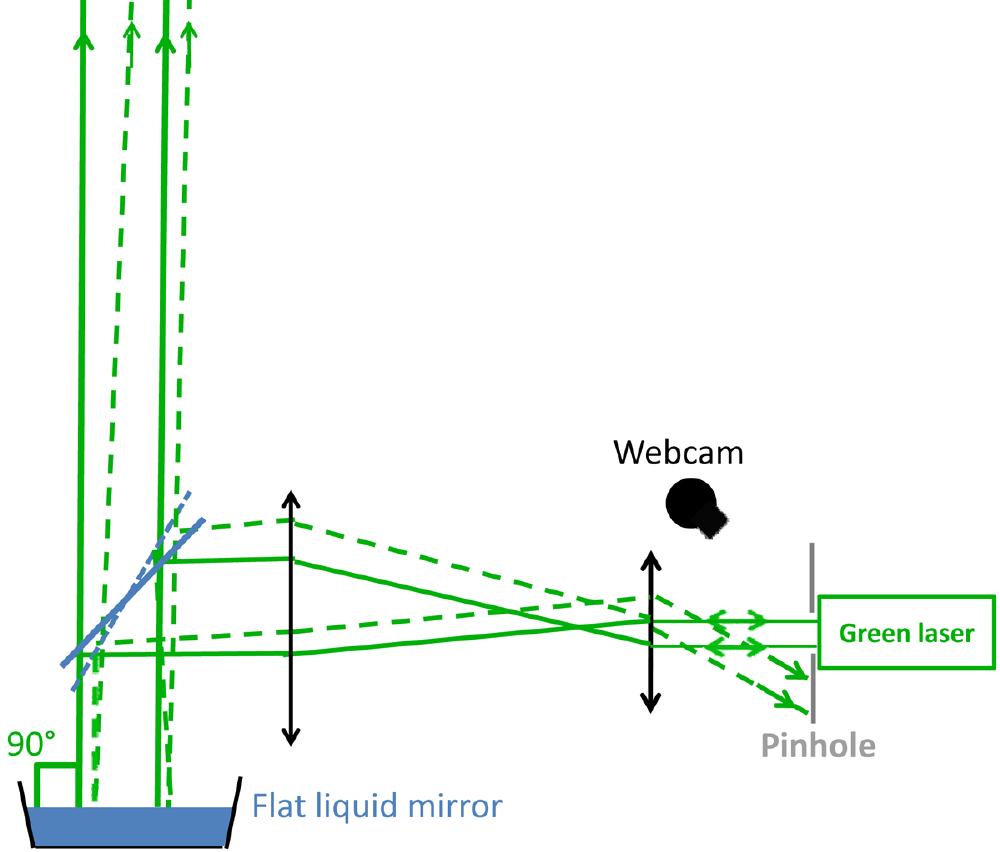}
  \caption{Principle for creating a vertical laser reference: a pellicle beam splitter above a flat liquid mirror is
    tilted in a such a way that the green laser goes back through a
    pinhole filmed by a webcam i.e. the fraction of the green laser
   light transmitted is perfectly vertical and can be used as a
   reference to align the metrology gondola above the $C'$ point (Fig. \ref{tableOptiquephoto}) of
  the metrology table. The dashed lines show the light rays if the beam splitter is misaligned. The reached accuracy is within one to two arcsec.}
 \label{laservertical}
\end{figure}

\noindent To optimize the system, it would be interesting to measure the closed loop bandwidth of the system (rejection function). To achieve that, it would be necessary to implement a recording of the PSD signals and of the correction sent to the motors in the servo loop software. Those data should be considered in
relation with recorded wind speed measurements. Moreover, several anemometers at different levels would probably be necessary, as the wind speed distribution changes significantly in different wind systems: wind gust, wind from N, S, partly cloudy, etc. We have already observed a wind of 10 $km/h$ at the level of the balloon and practically zero at the ground. Moreover, the oscillation frequencies will probably change with the focal gondola payload (not installed yet). As described at the top of this section, the servo loop works at $10\,\hertz$ (see appendix \ref{cable}). It could be interesting to optimize it . In the future, such a wind dependent full optimization could be
justified for a large scale diluted telescope project. However, such a
detailed characterization goes beyond the scope of this paper. Here, we demonstrate that the servo loop system is largely good enough to get stable metrology fringes under 20 $km/h$ wind conditions. 

\begin{figure}[h]
  \centering
  \includegraphics[width=8 cm]{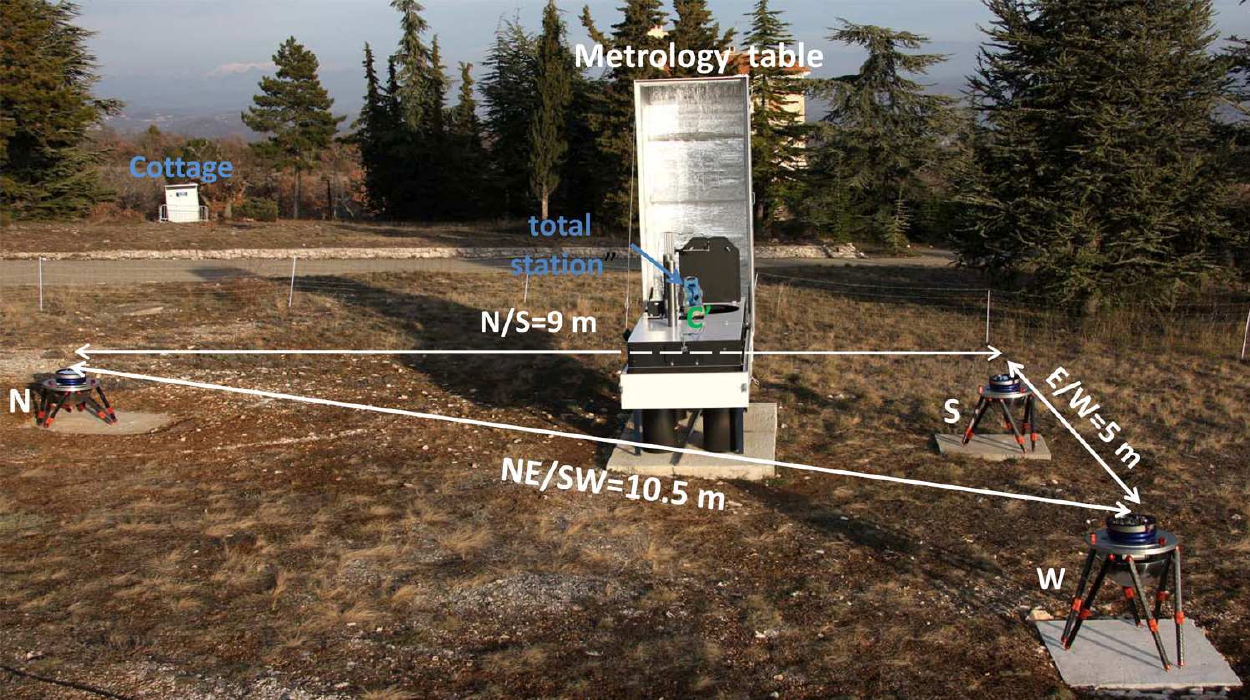}
  \caption{Photo of the Carlina mirrors. We see the three baselines of
    $9\,\meter$, $5\,\meter$, and $10.5 \,\meter$ oriented respectively N/S,
    E/W and NE/SW, around the metrology table described in
    Fig.~\ref{tableOptiquephoto}. The black hood is closed, and the
    trap door of the field mirror is open as during the observations.
    The total station helps to position the primary mirrors before observation. On this
    picture, the total station points to a corner cube positioned
    on the edge of the W mirror (Fig.~\ref{hexapod}).}
  \label{photoMirroirTable}
\end{figure}

\subsubsection{Alignment of the primary mirrors}
\label{alignmirror}

The total station is placed in the middle of the metrology table (point
$C'$ in Fig.~\ref{tableOptiquephoto}).
 It is used to point a
corner cube (centered on a sphere within $\approx 10\micro\meter$) that is
placed successively at three points on the edges of each mirror (Figs. \ref{hexapod}, \ref{photoMirroirTable}).
Two manual micrometer screws allow adjustment of the tip-tilt.  The piston is
adjusted by a micrometric vertical translation stage.

\subsubsection{Alignment of the primary mirrors with one micron accuracy}\label{onmicron}

We used the LEUKOS-SM (class 3b) supercontinuum laser source (Fig. \ref{tableOptiquephoto}). 
We adjusted the tip-tilt of each primary mirror in order to superpose the
three spots (images of the supercontinuum source created by each
primary mirror) on the field mirror
(Figs.~\ref{shematMetrology},~\ref{simuzemax},
and~\ref{tableOptiquephoto}). Then, we searched for the fringes in
$25\micro\meter$ steps, first using the vertical translation stage of
the W mirror (piston screw in Fig. \ref{hexapod}). Finally, the piston of the N mirror was adjusted
to find the N-S and the N-W fringes. Fringes have been
detected by scanning less than $200\, \micro\meter$. It shows that the
total station is enough accurate  to ensure a good prealignment. The total
stations and distance meter technologies are in constant
progress. In the future, by replacing the supercontinuum source by a
two-mode laser telemetry (Courde et al. \cite{Courde}; Courde et
al. \cite{Courde2}), it could be possible to position the 
primary mirrors within one micron without searching for the white
fringes (present technology).

\noindent The piston of the W and N mirrors are finely
adjusted to center the intersection of the three set of fringes close
to the middle of the speckle image (Fig.~\ref{franges}). It is remarkable that Fig.~\ref{franges} looks like simulated fringes (Fig. \ref{simuzemax}) but with additional turbulence (speckles). 
\noindent The same procedure can be used to adjust the tip-tilt and
piston of any additional mirror, by centering the fringes in the
speckle image, and at the crossing point of all the fringes (Fig. \ref{franges4mirroirs}).

\begin{figure}[h]
  \centering
  \includegraphics[width=8 cm]{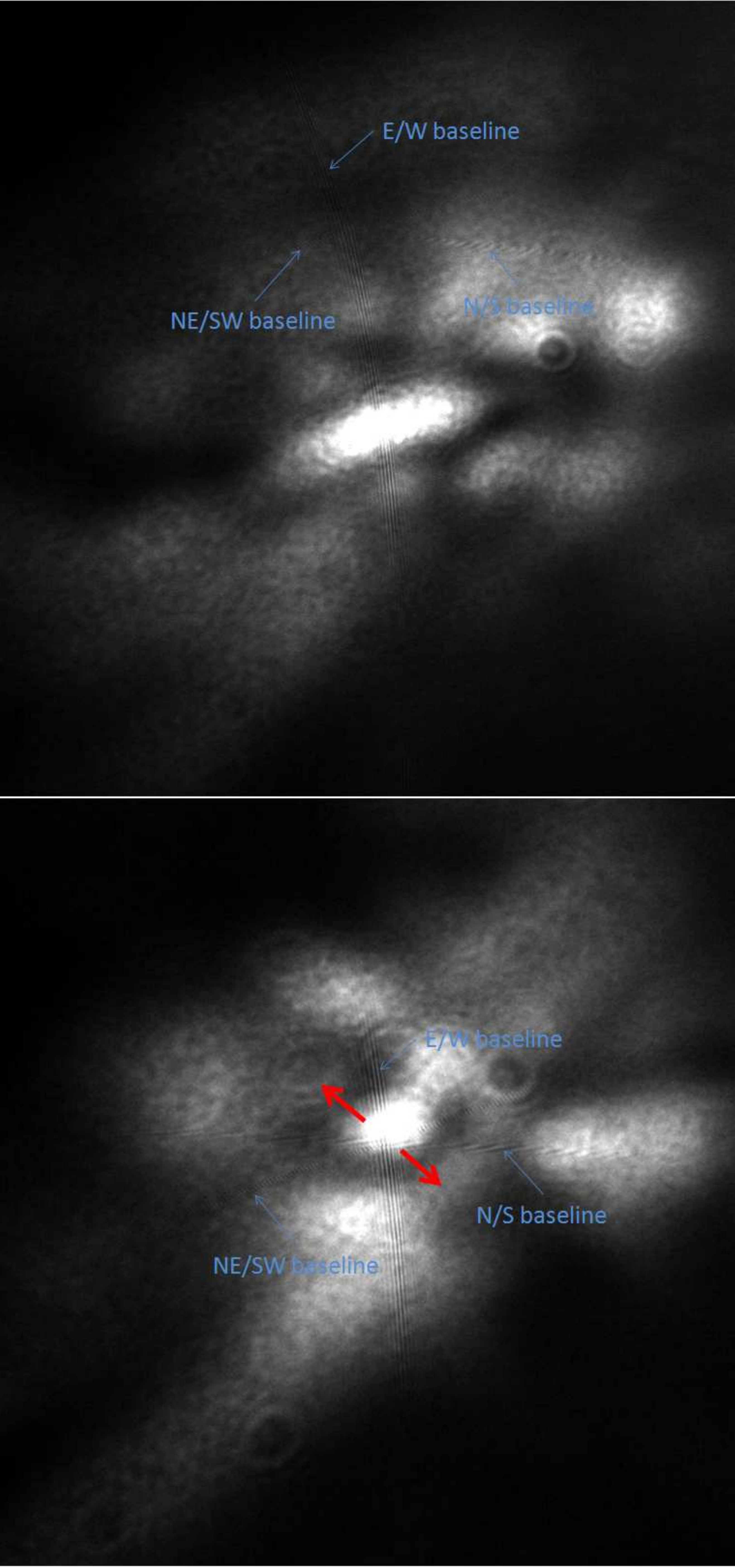}
  \caption{Fringes from the three baselines. Top, $50\, \micro\meter$ piston
    error on the N primary mirror. Bottom, fringes are centered:
    the three mirrors are cospheric  within a few microns (not
    possible to have a better adjustment due to atmospheric
    turbulence).  Fringes are subsampled on the pixels of the CCD for
    larger baselines (N/S, NE/SW) and they appear with a moire
    pattern. To avoid the moire, fringes are sufficiently stable to be
    magnified by a factor 10 (using a lens in front of the CCD) but
    the speckle image would become larger than the CCD. We have
    checked that the moire pattern does not disturb the accuracy needed to center the fringes (Fig. \ref{fringestability}). 
    The red arrows indicate the dominant direction of the oscillations of the fringes on the CCD. }
  \label{franges}
\end{figure}

\section{Results}\label{fringes}

The optomechanical  systems described in this paper has been tested following four main steps:\\

\begin{itemize}
\item Alignment of the metrology gondola (cables length adjustment, test of the vertical green laser, etc.)
\item Servo loop (interaction matrix creation, PID parameters, bug corrections, etc.)
\item Primary mirrors alignment and fringes detection
\item Characterization of the system stability
\end{itemize}

\noindent A log of the completed stages is provided in the appendix \ref{log}.
Beside the 380 fringe images (5x76 frames recorded on February 23,
2011) to characterize the system stability (as shown
in Figs. \ref{fringestability} and \ref{franges15min}), we also recorded other data. However, these additional
fringe frames don't constitute a coherent data set, and we therefore
don't show these results here. Among the numerous measurements performed
with the prototype, we mention fringe imaging with only one baseline, piston adjustment of the N mirror (see e.g. Fig. \ref{franges}), fringe
measurements after interruption and restart of the servo loop, etc. In
total, 21x76 = 1596 images were recorded without losing fringes during tests
lasting for four hours.

\subsection{System stability}\label{systemstabi}


\noindent In order to study the system's stability, we have plotted the fringe
positions on the metrology camera, as a function of time
(Fig.~\ref{fringestability}). First, we tried to observe the fringe
motions relative to the diffraction envelopes images of the subapertures. However,
we realized that the photocenter of the diffraction envelopes moves
on the camera much more than the fringes. This means that the diffraction envelope photocenter
displacement is dominated by the atmosphere (tip-tilts and speckle
effects) and not by the girder gondola oscillations. Indeed, the fringes and
the diffraction envelopes all move together with the gondola
oscillations, while the atmospheric tip-tilt on a subaperture moves
only the photocenter diffraction envelope (not the fringes) on the CCD.

\noindent We then measured the fringe displacements relative to the
camera pixels. Figure \ref{fringestability} shows that the fringes
oscillate around a fixed position during $\approx 7\second$, the time needed
to fill the PC memory ram (76 exposures at $11.3$ images per second over
$2048\times2048$ pixels). Fringe residual motions are due to the metrology gondola residual oscillations and the
atmospheric piston. In Fig.~\ref{fringestability}, we see that the
maximum oscillations are about equal to $\pm220$ microns ($\pm30$ pixels). The fringes oscillate in a dominant direction (Fig. \ref{franges}), probably linked with the wind direction, showing that their motions are correlated well  with a metrology gondola oscillation. If their motions were dominated by the turbulence (differential piston on the primary mirrors), their displacement on the CCD would be in random directions.  This rectilinear oscillation explains  that the three curves (Fig.~\ref{fringestability}) look very similar in shape and amplitude. These three curves can be interpreted as the projection of a nonuniform rectilinear motion of a point (the intersection of the three fringes) on three axes: the axes perpendicular to the fringes, i.e in the direction of the baselines. By definition, a rectilinear motion of a point projected on any axis gives the same curve in shape on each axis. The three curves of Fig. \ref{fringestability} have about the same amplitude because the fringes oscillate in a direction very different from all the projection axes (see fig. \ref{franges}).\\

\noindent The results of the study of the system stability over a longer duration (15 min) are presented in Fig. \ref{franges15min}. The position of the fringes are averaged over seven seconds at each point in the lower panel of Fig. \ref{franges15min} and then the motion of the fringe center is only due to the mechanics (turbulence averaged in seven seconds). The fringes have moved by $\approx 120$ pixels ($ \approx 0.9 $ mm) during 15 min (Fig. \ref{franges15min}).  With the Zemax software, we evaluated that a displacement of the fringes of $120$ pixels on the metrology CCD corresponds to a motion of $\Delta_{\Omega(x,y)}=250\,\mu m$ of the top of the metrology gondola; i.e., we stabilized the center of curvature of the metrology mirror within $250\,\mu m$ during 15 min! This is very satisfactory considering that some of optical components of the system are suspended under a balloon. The optics (field mirror, CCD size, etc.) were dimensioned to tolerate 16 times larger displacements  ($\Delta = \pm 2$ mm as given in the specifications of Table \ref{tablspec}). In fact, we never lost the fringes for four hours starting from the moment that we found them on the CCD.\\

\begin{figure}[h]
  \centering
  \includegraphics[width=8cm]{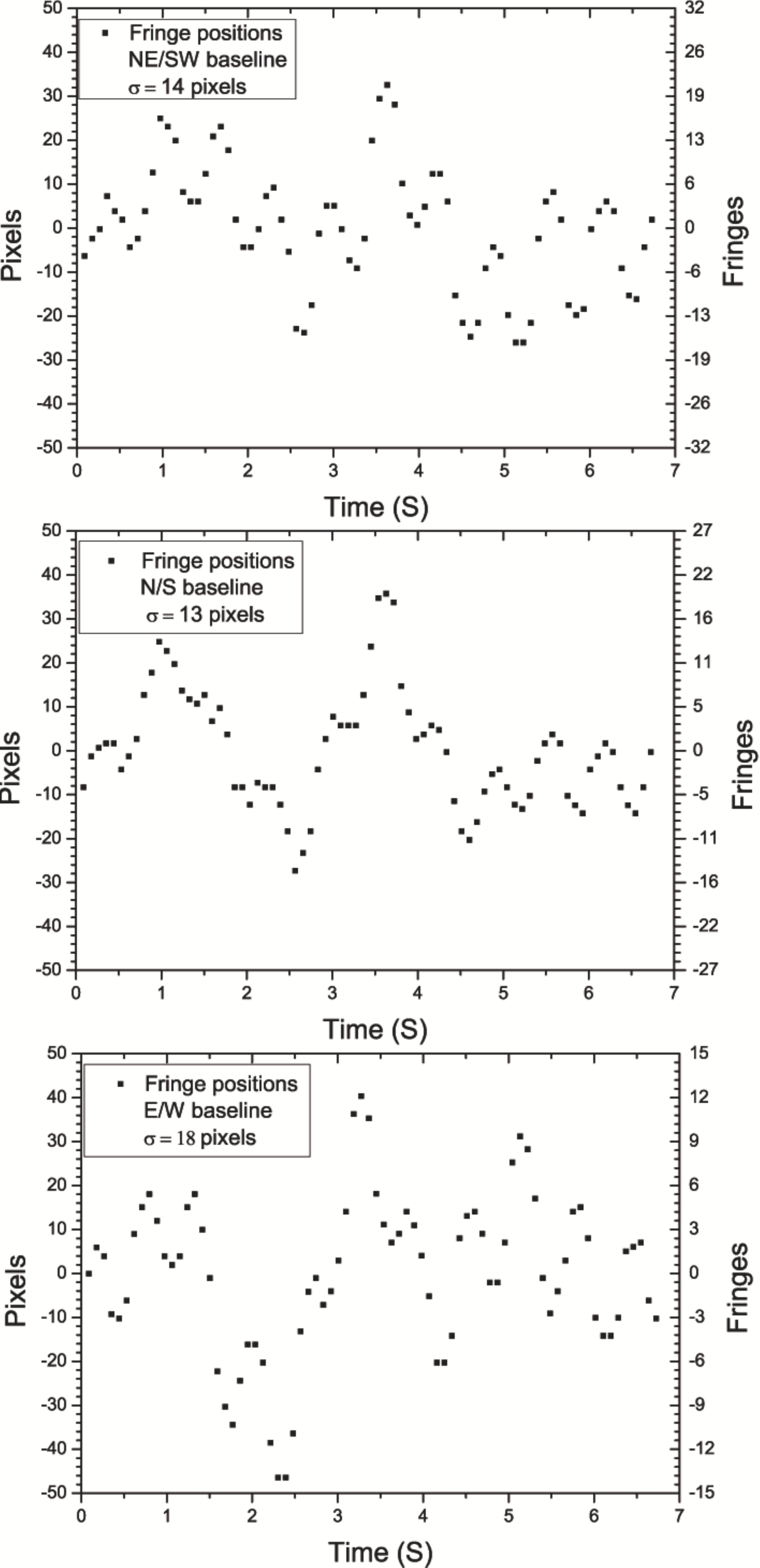}
  \caption{Displacement of the fringes (in the direction perpendicular
    to the fringes) from the three NE/SW, N/S and E/W baselines
    (Fig.~\ref{photoMirroirTable}). On the left, scales are in pixel; on the 
    right, in fringes. One fringe equals $1.56$, $1.82$, and
    $3.28$ pixels, respectively, for the  NE/SW, N/S, and E/W baselines.}
  \label{fringestability}
\end{figure}
\begin{figure}[h]
  \centering
  \includegraphics[width=8cm]{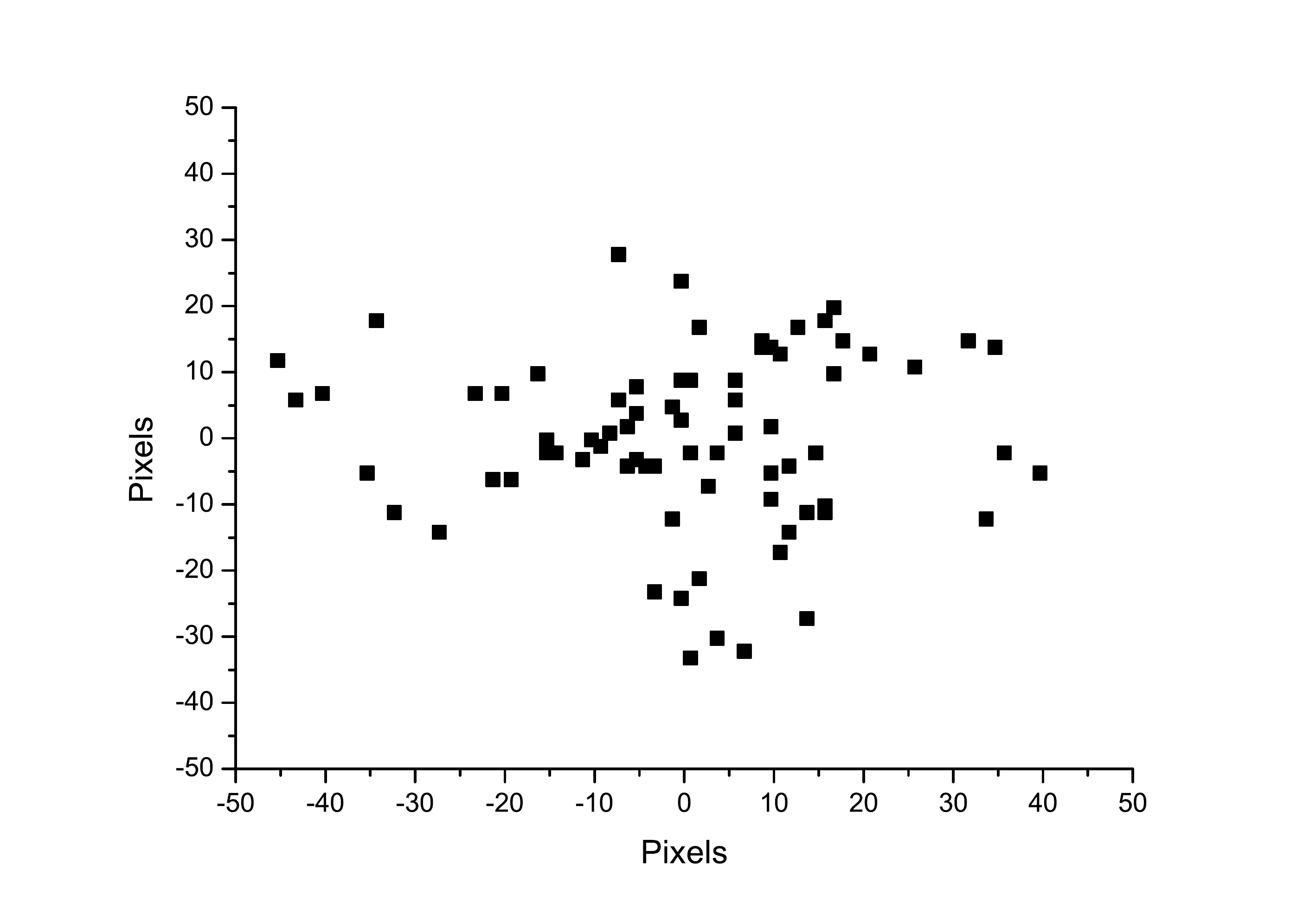}
  \includegraphics[width=8cm]{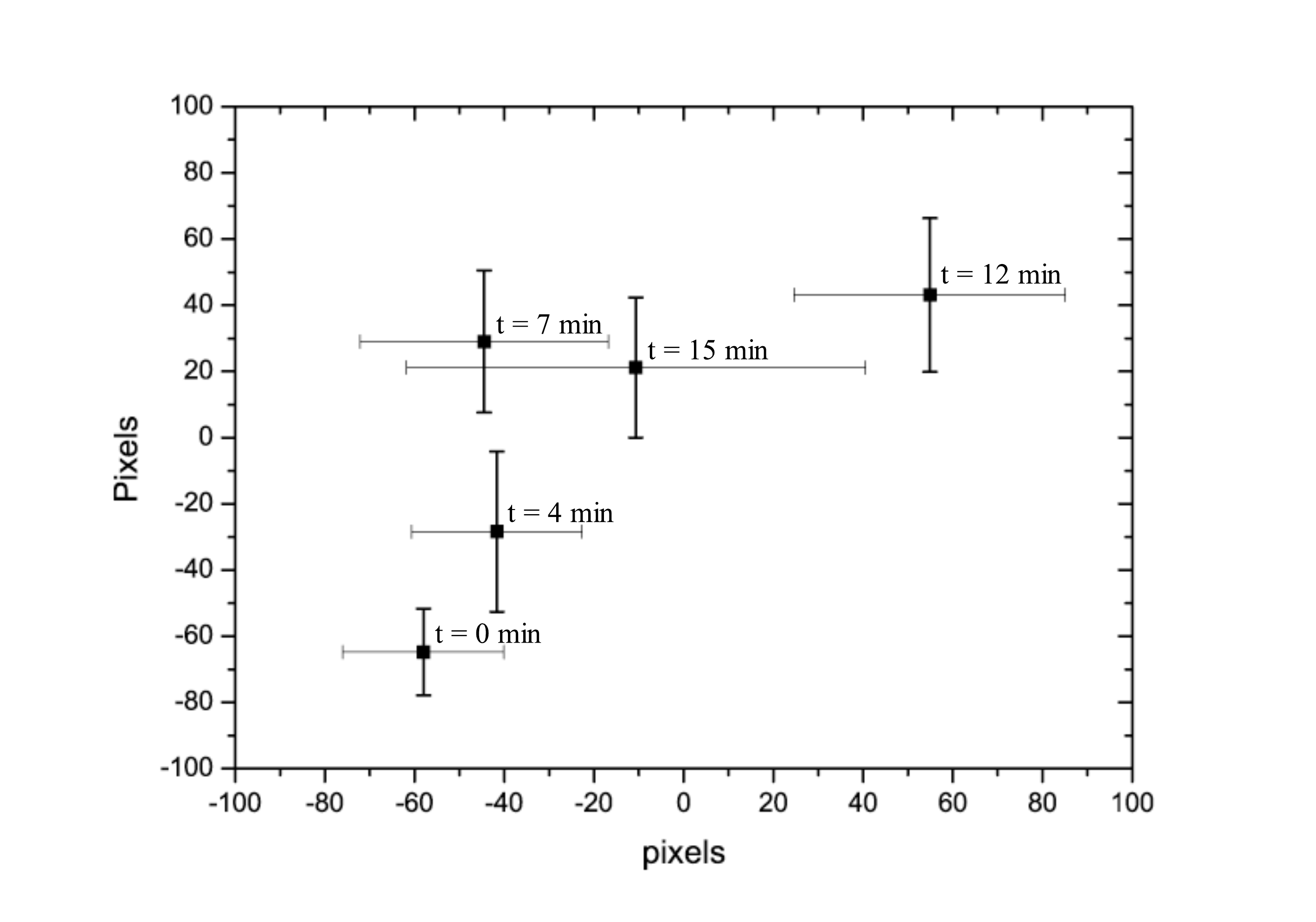}
  \caption{Top: Position of the fringes center (intersection of the three fringes systems) on the CCD during $\approx 7\second$ (76 exposures of 1 millisecond). Bottom: Each point is the mean position of the fringes center over 7 seconds (ex: the point at t=0 min is the mean of the top graph). The data have been recorded the 23/02/2011 during 15 min (see log Appendix \ref{log}).}
  \label{franges15min}
\end{figure}

\noindent Moreover, the fringes remains centered on the envelope (subapertures image) for 15 minutes at least, without touching the primary mirrors. This demonstrates the excellent stability of the mechanical supports, as well as the stability of the ground itself i.e. fringes are centered in the envelope with a few fringes accuracy. This result is well in the specifications of Table \ref{tablspec}.\\
\noindent For a scientific diluted telescope, the tens (or hundreds) of  mirrors will be motorized
and it will be easy to track the possible slow drifts. We will have to develop an algorithm
adapted to track the mean position of the fringes. A cross correlation of the fringes could be a good solution to explore this algorithm. We could call this new fringe sensor for diluted pupils: the CCFFS, for cross correlation Fizeau fringe sensor. 

\noindent To conclude, we have centered the fringes on a mean position in the envelope of diffraction of the subapertures (accuracy of a few fringes). Note that our
metrology is equivalent to an active system and cannot be used as
adaptive optics because the metrology laser beams do not follow the same
optical path as the star light (see Sect. \ref{noncommon}). Adaptive optics will have to be
implemented in the focal gondola (could use a Fizeau fringe sensor, such as the CCFFS proposed in this paper). Then, it will be possible to make longtime
  exposures in order to increase the sensitivity.\\
  \noindent We have demonstrated that it is possible to align the primary mirrors with a metrology system attached at the center of curvature under a captive helium balloon.
This is an important step toward demonstrating the possibility of building such an interferometer
using a spherical diluted primary mirror. In Appendix \ref{discussion}, we discuss how
to search for sites where such a project could be built and how to define the array of primary mirrors (number of mirrors, pupil redundant or not, etc.) depending the science goals. Finally, we propose to build a scientific demonstrator that we will call the LDT. Theoretically, an
LDT working with 100 subapertures of  $25 \centi\meter$ diameter
should be able to observe objects as faint as $m_v=15$ (Le Coroller et
al. \cite{lecoroller2}; De Becker et al. \cite{DeBecker}). The
limiting magnitude as a function of the mirror surface will be
determined experimentally when our prototype is operating with its
focal gondola. 


\section{Conclusions} \label{conclusion}

The main result of this work is the demonstration of the capabilities and performances of an original metrology system designed for aligning the primary elements of a diluted spherical mirror under a helium balloon.
This is an important step that demonstrates the
feasibility of an interferometer with a spherical diluted
mirror. Active optics is also an important
step toward implementing adaptive optics.\\

\noindent We have developed the servo loop and the metrology to cospherize three mirrors with a spacing of approximately $10\,\meter$
on a $71.2\, \meter$ curvature radius sphere. The servo loop stabilizes a metrology mirror (near the curvature center of the diluted primary mirror $\approx 71 \meter$ above the ground) with an accuracy better than $5\,
\milli\meter$, and within 250 microns for its curvature center.  The optomechanical  design also allows putting heavy metrology devices (lasers, camera, optics, etc.) on a table at ground level. The metrology fringes are observed behind a field mirror in a pseudo-pupil
plane where they are stable. We have fully responded to the specifications of Table \ref{tablspec}. In particular:

\begin{itemize}
\item Stability of the mounts that carry the primary mirrors $\approx 1 \mu$m
\item Stability of the curvature center of the metrology mirror under the helium balloon, demonstrated in this paper over 15 min : $\Delta_{\Omega(x,y)}=250\,\mu m$ and $\Delta_z<5\,mm$
\end{itemize}

\noindent To our knowledge, this is the first time that a
supercontinuum laser source is used in order to equalize the optical
paths of an astronomical interferometer. Such an intense source
provides fringes with a high $S/N$ ratio that allows the primary segments to be positioned very
accurately, down to the seeing limited value. This was made possible by prepositioning the segments using a total station.
We note that future laser telemetry technologies (Courde et
al. \cite{Courde}) could be alternative solutions to position the
mirrors of diluted telescopes or interferometer delay lines.\\

\noindent In a future article, we will report on observations made with
a focal gondola currently under construction.  We will test the
optical train from end to end (Mertz Corrector,
pupil densifier, etc.)  and will clarify experimentally the limiting
magnitude.  We will measure the stability of this gondola under the
stabilized tripod (described in this article). Carlina with the MMT (Carleton \cite{Carleton}) and LBT (Kim
et al. \cite{Kim}) could constitute a new family of telescopes called
diluted telescopes. We propose to build in the next 10 years a
scientific demonstrator with an aperture of $\approx 100\,\meter$ that we will
call the LDT. In a second stage, a Very
LDT could be installed in an area with extremely
good seeing conditions.\\

\noindent Because there are no simple solution, the studies on the
compact arrays of mirrors positioned on a sphere (as proposed here) or linked by delay lines have to be pursued. If delay lines are
used, the light beams will have to exit the telescopes with a maximum
of two to three reflections using, for example, an ``Alt-Alt'' mount. To minimize
the absorption due to many reflections on numerous mirrors, the light
could be transported in single mode fibers (Perrin \cite{Perrin}).
More studies are required for determining the characteristics
of the atmospheric turbulence in valleys compatible with such projects:
located relatively high up  in the mountains, in a valley oriented in the E-W
direction, and with a nearly hemicylindrical shape (appendix \ref{discussion}).  If no natural
valleys with good seeing conditions can be found, other solutions
using pylons could be explored.

\begin{acknowledgements}
  This research has been funded by CNRS/INSU and Coll\`ege de
  France. Mechanical elements were built by the technical group at OHP
  and Nice observatory.  We thank Chris Bee for improving the English.
   We are grateful to Jean Surdej, Michel Lintz, and Luc Arnold for some
  useful corrections and suggestions.  We are grateful to student
  and to the people who helped us during the long nights of tests: Romain
  Pascal, Jean-Philippe Orts, and Julien Chombart.  We are grateful to the
  LEUKOS company that helped us to use their supercontinuum laser.
  Google earth and IGN have been used to find sites adapted to a
  spherical diluted telescope.
\end{acknowledgements}




\begin{appendix} 
\section{Carlina specifications}
  \label{CarlinaSpec}
 In this appendix, the specifications of each part (primary mirror, focal gondola, metrology gondola and holding system) are provided (Table \ref{tablspec}).
\begin{center}
\begin{table*}[!t]
\begin{tabular}{|p{1.3cm}|p{3.7cm}|p{3.7cm}|p{3.7cm}|p{3.7cm}|}
  \hline
  \textbf{Parts} & \textbf{Characteristics} & \multicolumn{3}{c|}{\textbf{Specifications}}\\
  \cline{3-5}
   & & High Level & Low Level &  OHP prototype\\
   \hline
  Primary \newline mirror & Numerous spherical mirrors, all with the same curvature radius. & - Mirrors aligned on a sphere (coherencing) \newline - The numbers of mirrors and their positions have to be adapted to the science targets. & Piston and Tip-Tilt of each mirror must be controlled with an accuracy of one $\lambda$.& - 3 mirrors; R=71.2m \newline - Bases: 5m, 9m, 10.5 m \newline - Mounts stability and micrometric screws accuracy \newline $\approx 1 \mu m$ \\
  \hline
  Focal \newline  gondola & \underline{Used to obtain stellar images:} \newline Constituted with a spherical aberration corrector (Mertz), a guiding system, and a focal instrument (densifier, AO, etc.). & - Track the star on the focal sphere (R/2). \newline - The drift during the tracking has to be smaller than one fringe per exposure time. \newline - Have to be positioned at the correct focus. & - Maximum tracking velocity $7.3\times10^{-5}\,\frac{R}{2}$ \newline - Velocity drifts $< \lambda f/B$ per exposure time  \newline - Focus tolerance (spots shifted by half their size) \newline $\Delta F \approx  \pm(\frac{f}{B})(\frac{f}{r_o})\lambda$& - Maximum tracking velocity $=\,2.6\, mm/s$  \newline  - Maximum velocity drifts $=\, 1.7 \,mm/s$ for $1\, \milli\second$ exposure time\newline - Focus tolerance $=\,\pm 0.6\, mm$  \\
  \hline
  Metrology \newline  gondola & \underline{Used to align the primary} \newline \underline{ mirrors:} \newline The Tip-Tilt of each primary mirror is adjusted by superimposing all the return images of a source at the curvature center, while the piston is adjusted by searching the white fringe. \newline In practice, the source and the metrology CCD are placed on the ground and the metrology gondola is made of a convex mirror attached at the bottom of a girder (Fig. \ref{photoGondoGirder}). &  - The top of the girder gondola ($\Omega$) must be stable, in a such a way the return beam goes on the field mirror at the ground (Figs. \ref{shematMetrology}, \ref{OHPProto}).\newline - The altitude of the girder gondola must be accurate to focalize the metrology source near the center of the diluted primary mirror (also on the field mirror). & -  The maximum tolerable displacement ("horizontaly") of $\Omega$  is \newline \center{$\Delta_{\Omega(x,y)} =\frac{r_{field\, mirror}}{R}\frac{f_{metrology}}{2}$} \newline \newline - The position tolerance of the metrology gondola in altitude (vertical displacement) is : \center{$\Delta_z=\pm \frac{\lambda}{B}\frac{R^2}{r_o}$} & - $\Delta_{\Omega(x,y)} = \pm 2\,mm$ \newline - $\Delta_z = \pm 2.5\, mm$ \newline \newline with $r_o=100\,mm$, $r_{field\, mirror}=300\,mm$, $R=71.2\,m$ and $f_{metrology}=1m$  \\
  \hline
  Holding \newline  system & The Holding system carries the metrology gondola, and the focal gondola above the primary mirrors.   & It can be made of cables under an helium balloon, or cables attached between mountains, pylons, etc. & - At least at the altitude of the curvature radius $R$ of the primary mirror above the primary segments. \newline - It uses material with high tensile strength and Young's modulus: Kevelar, Carbon, etc. \newline - The holding system can be stabilized.
 & - At OHP, a helium balloon at 120 m altitude tightens a tripod of PBO cables. \newline - The top of this stabilized tripod is at $\approx 71\,\meter$ above the ground.
  \\
  \hline
\end{tabular}
\caption{\label{tablspec} Carlina specifications. R is the curvature radius and $f=R/2$ the focal length of the diluted primary mirror; $r_{field\, mirror}$ is the radius of the field mirror; $f_{metrology}$ the focal length of the metrology mirror; $r_0$, the atmospheric Fried parameter ($\approx 100$ mm at OHP); B the baseline; the wavelength $\lambda=0.5\,\mu m$}
\end{table*}
\end{center}

\end{appendix}


\begin{appendix} 
\section{The Kevlar cable response}
\label{cable}
In this appendix, we briefly describe the approach for selecting the
frequency of the servo loop to stabilize the metrology gondola.
At this stage in characterizing the system, we focused on the
response time of the kevlar cables. We horizontally hung a 100 m kevlar
cable at about 1 meter above the ground. At one end, the cable passes
through a pulley and is attached to $17\,\kilogram$ mass, exerting therefore a
tension on the cable that is similar to that of the tripod of the
prototype. The other end of the cable is attached to a motorized winch.
We used a linear position sensor to detect the mass displacement.\\
\newline The Fig. \ref{figcable} shows the open-loop step response of the system to a unit movement. The rising time (time delay between the activation of
the motorized winch and the detection of the mass motion) is $\approx 50 \, \milli\second$. This time interval constitutes the lower limit on the time reaction of the servo loop. In such circumstances, a reasonable period for the
operation of the loop could be selected as twice the lower limit imposed
by the cable reaction. We therefore set the loop frequency to $10\,\hertz$. 
The pseudo period and the settling time observed here (Fig. \ref{figcable}) are not significant, because they mainly depend on the mass and on the geometry of the system. Indeed, $17\,\kilogram$ is the tension created by the balloon in the cables, but the mass of the metrology gondola is much lighter $\approx \, 3\, \kilogram$. This experiment was only destined to evaluate the response time and not the resonant and damping frequency. The frequency of $10\,\hertz$ will be optimized by better characterizing the response of the full system (cable tripod with metrology and focal gondolas) with the PBO cables rather than the response of a single tensioned Kevlar cable (see also discussion Sect. \ref{servoloop}). The rising time of the PBO cables could also be faster than Kevlar because their Young modulus is four times higher.

\begin{figure}[!h]
    \centering
    \includegraphics[width=8.8cm]{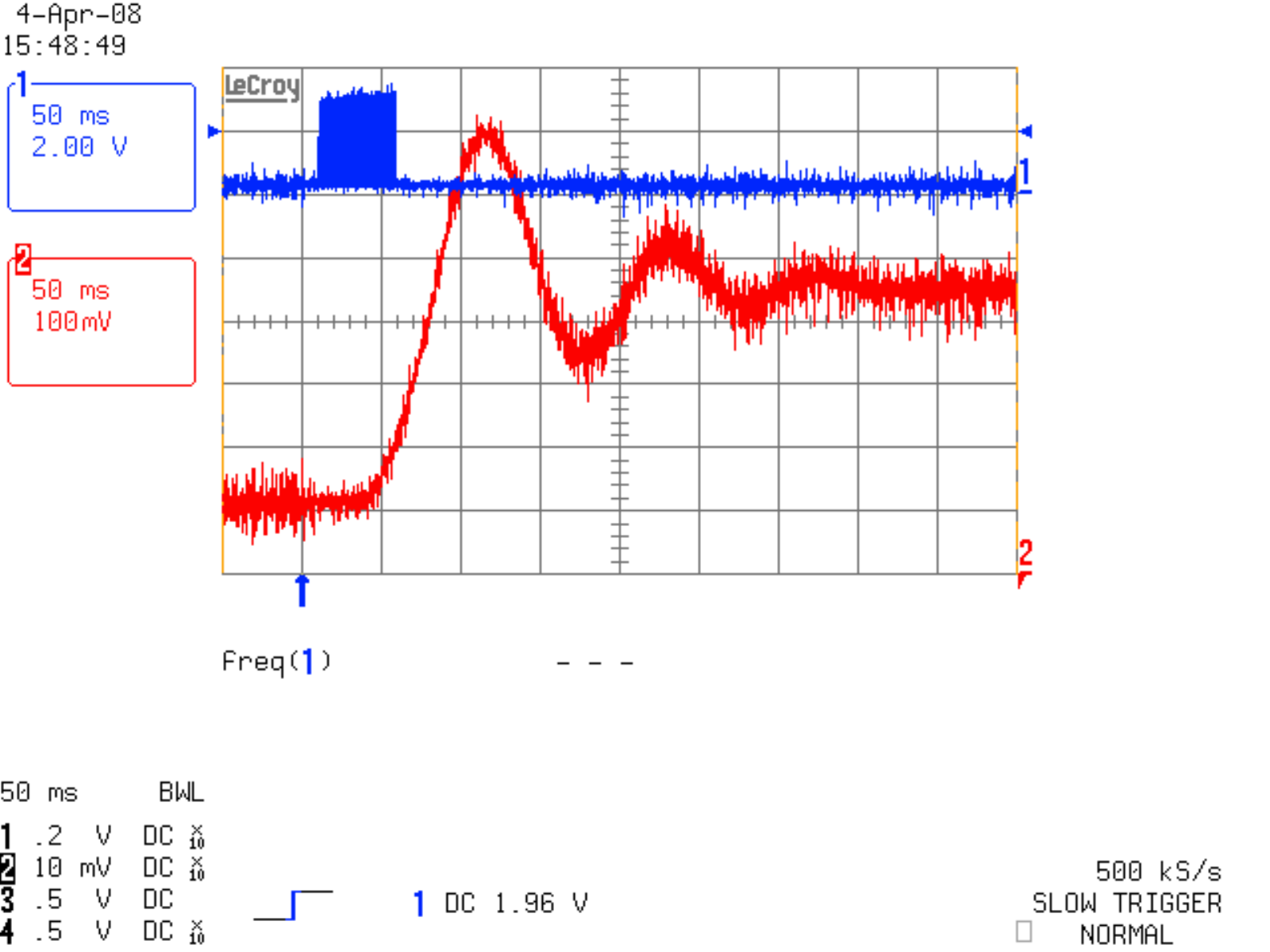}
    \caption{Oscilloscope window: Open-loop step  response of a single tensioned Kevlar cable. The blue curve shows the speed of the motorized winch, actuated for $50\, \milli\second$ (horizontal scale is $50\, \milli\second$/square). The red curve shows the observed position of the weight versus time. The rise time of this experimental setup is $\approx \, 50\, \milli\second$, followed by three dumped oscillations. Settling time is $300\, \milli\second$.}
    \label{figcable}
  \end{figure}
\end{appendix}


\begin{appendix}
\section{Experiment log} \label{log}
Each night of test represents about 20 hours of work to install the balloon, cables, etc. We divided the alignments in four main tasks. Each task has been completed in one to three nights (see Table \ref{tabllog}). We always worked in a low wind ($<20$ km/h measured with an anemometer attached to the helium balloon). Above $20$ km/h windspeed, the experiment went down (a cable of the tripod is slackened), and it was not possible to work even with a servo loop. \\

\noindent All the data were obtained on February 23, 2011. We recorded 1596 images of fringes distributed over four hours. We present in this paper the data extracted from 380 images obtain during 15 min in a coherent block (without touching anything).

\begin{table*}[!t]

\begin{tabularx}{10cm}{|X|p{1.5cm}|p{2cm}|}
\hline
\textbf{Tasks} &  \textbf{dates} & \textbf{Sect.} \\
\hline
Alignment of the metrology gondola (cables length adjustment, test of the vertical green laser, etc.) & 27/07/2009 \newline 3/12/2009 \newline 7/02/2010  &  \ref{alignmetrogondol} \\
\hline
Servo loop (interaction matrix creation, PID parameters, bug corrections, etc.) & 08/03/2010 \newline 14/04/2010 & \ref{servoloop} and \ref{systemcalib}\\
\hline
Primary mirrors alignment and fringes detection & 26/07/2010 \newline 09/2010  & \ref{alignmirror} and \ref{onmicron}\\
\hline
Characterization of the system stability; \newline fringe image recording (Figs. \ref{fringestability} and \ref{franges15min})& 23/02/2011 & \ref{fringes} and \ref{systemstabi}\\
\hline
\end{tabularx}
\caption{Log of the main runs. The dates correspond to the beginning of the night.}
 \label{tabllog}
\end{table*}

\end{appendix}


\begin{appendix}  

  \section{System calibration}
  \label{systemcalib}

  The displacements of the photocenters (image of the corner cube) on
  both PSDs are linked to the length variation of the tripod cables
  by the linear equation:
  \begin{equation}
    \Delta \vec{P}=\pmatrix{\Delta P_{x1}\cr\
      \Delta P_{y1}\cr
      \Delta P_{x2}\cr
      \Delta p_{y2}\cr}=A\cdot\Delta\vec{L}=
    \pmatrix{
      {x'_1}&{x''_1}&{x'''_1}\cr
      {y'_1}&{y''_1}&{y'''_1}\cr
      {x'_2}&{x''_2}&{x'''_2}\cr
      {y'_2}&{y''_2}&{y'''_2}\cr
    }\pmatrix{
      \Delta L_1\cr
      \Delta L_2\cr
      \Delta L_3\cr},
    \label{eqAsserv}
  \end{equation}
  where $\Delta \vec{P}= \vec{P_{measured}}-\vec{P_{reference}}$;
  ($\Delta P_{x1}$, $\Delta P_{y1}$) and ($\Delta P_{x2}$, $\Delta
  P_{y2}$) are the positions of the photocenters with respect to a
  reference point (noted $\vec{P_{reference}}$), respectively on PSD1,
  and PSD2.  $\Delta L_j$ is the length variation of the tripod cable (Fig. \ref{photoGondoGirder});
  $A$ is an interaction matrix determined using the following
  process: ideally, the system must be calibrated during a night
  without wind and oscillations of the tripod.  In practice, the servo
  loop works well even if the interaction matrix has been created
  during a slightly windy night; the
  system returns the girder gondola, in a stepwise linear fashion
  toward a reference point.
  In order
  to determine the interaction matrix values, we actuate successively
  each motorized winch.

  \noindent The MW1 motorized winch (Fig.~\ref{shematCarlinaProto})
  unwinds the cable by a unit value $\Delta L_1$ without moving
  $MW2$ and $MW3$. Then, the coordinates of the photocenters on each PSD
  gives a vector ($\Delta P_{x1}$, $\Delta P_{y1}$, $\Delta P_{x2}$,
  $\Delta P_{y2}$) equal to the first column of $A$, the interaction
  matrix by replacing $\Delta L_1=1$, $\Delta L_2=\Delta L_3=0$ in
  equation \ref{eqAsserv}:
  \begin{equation}
    \pmatrix{
      {x'_1}&{x''_1}&{x'''_1}\cr
      {y'_1}&{y''_1}&{y'''_1}\cr
      {x'_2}&{x''_2}&{x'''_2}\cr
      {y'_2}&{y''_2}&{y'''_2}\cr
    }\pmatrix{
      1\cr
      0\cr
      0\cr}=\pmatrix{{x'_1}\cr
      {y'_1}\cr
      {x'_2}\cr
      {y'_2}\cr}=\pmatrix{\Delta P_{x1}\cr
      \Delta P_{y1}\cr
      \Delta P_{x2}\cr
      \Delta p_{y2}\cr}_{measured\, by\, MW1}
     \nonumber
  \end{equation}

  \noindent The other columns of the interaction matrix are found by
  actuating the second and third motor winches:
  \begin{equation}
    \pmatrix{
      {x'_1}&{x''_1}&{x'''_1}\cr
      {y'_1}&{y''_1}&{y'''_1}\cr
      {x'_2}&{x''_2}&{x'''_2}\cr
      {y'_2}&{y''_2}&{y'''_2}\cr
    }\pmatrix{
      0\cr
      1\cr
      0\cr}=\pmatrix{{x''_1}\cr
      {y''_1}\cr
      {x''_2}\cr
      {y''_2}\cr}=\pmatrix{\Delta P_{x1}\cr
      \Delta P_{y1}\cr
      \Delta P_{x2}\cr
      \Delta p_{y2}\cr}_{measured\, by\, MW2} \nonumber
  \end{equation}
  and
  \begin{equation}
    \pmatrix{
      {x'_1}&{x''_1}&{x'''_1}\cr
      {y'_1}&{y''_1}&{y'''_1}\cr
      {x'_2}&{x''_2}&{x'''_2}\cr
      {y'_2}&{y''_2}&{y'''_2}\cr
    }\pmatrix{
      0\cr
      0\cr
      1\cr}=\pmatrix{{x'''_1}\cr
      {y'''_1}\cr
      {x'''_2}\cr
      {y'''_2}\cr}=\pmatrix{\Delta P_{x1}\cr
      \Delta P_{y1}\cr
      \Delta P_{x2}\cr
      \Delta p_{y2}\cr}_{measured\, by\, MW3}
     \nonumber
  \end{equation}

  \noindent The inverse problem must be solved to determine how much
  each motor has to move to return the girder gondola to its initial
  position: $\Delta\vec{L}=M\cdot \Delta \vec{P}$.

  \noindent The inverse $A$ matrix is calculated by using the
  Singular Value Decomposition (SVD) method for the rectangular
  matrices (Jacubowiez et al. \cite{Jacubowiez}). The matrix $M$ is
  computed using the Matlab software SVD function.  Finally, a field
  programmable gate array (FPGA) computes $\Delta\vec{L}$ in
  real time:
  \begin{equation}
    \pmatrix{
      \Delta L_{1}\cr
      \Delta L_{2}\cr
      \Delta L_{3}\cr}=
    \pmatrix{
      m_{11}&m_{12}&m_{13}&m_{14}\cr
      m_{21}&m_{22}&m_{23}&m_{24}\cr
      m_{31}&m_{32}&m_{33}&m_{34}\cr}
    \pmatrix{
      \Delta P_{x1}\cr
      \Delta P_{y1}\cr
      \Delta P_{x2}\cr
      \Delta P_{y2}\cr}
    . \nonumber
  \end{equation}

  \noindent The servo loop system is piloted via a graphical interface
  (Fig.~\ref{Windasserv}).

  \begin{figure}[h]
    \centering
    \includegraphics[width=8.8cm]{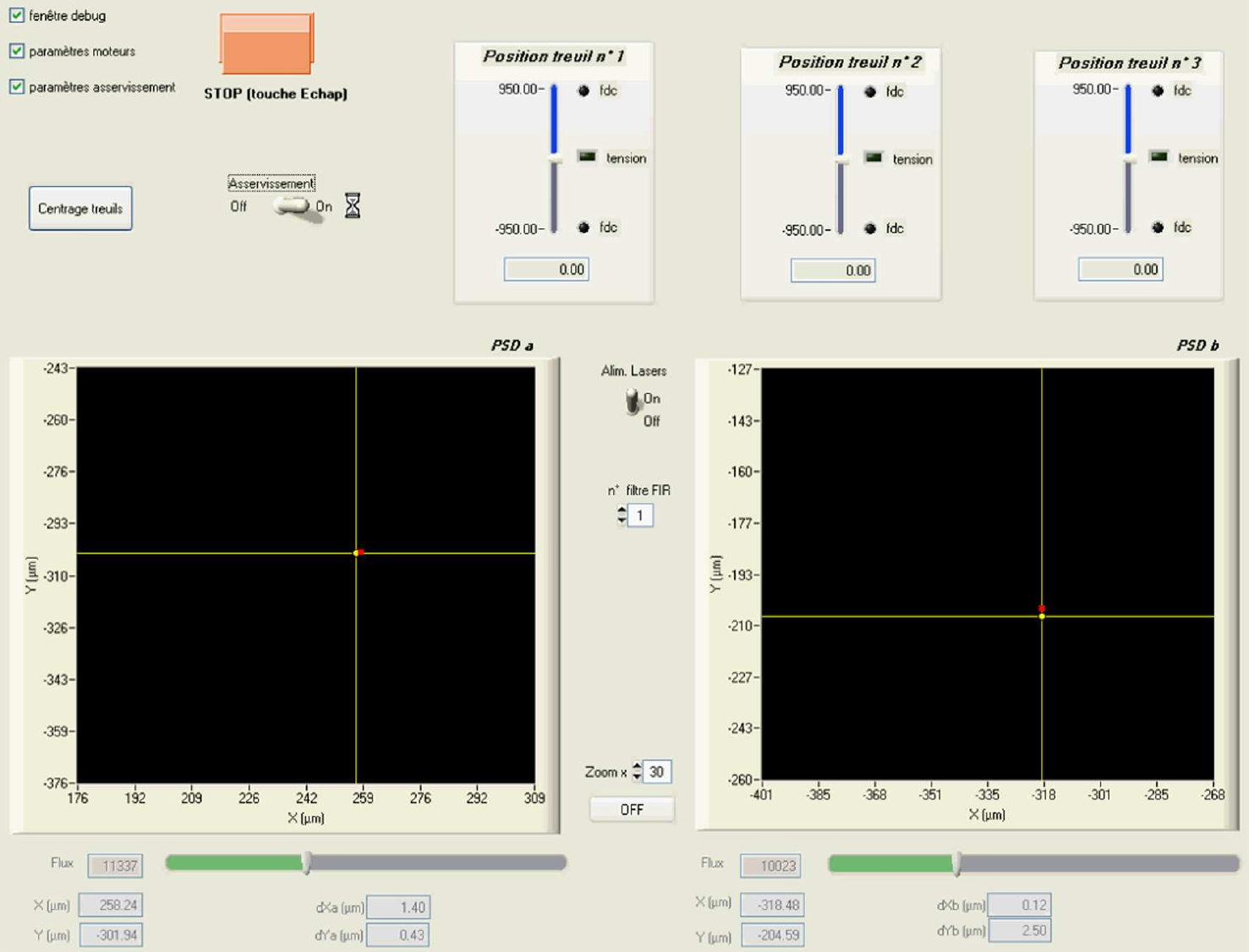}
    \caption{Control window of the servo loop. The horizontal green
      bar shows the flux on the PSDs (it becomes red if the flux is
      too low). The red point is the position of the photocenter on
      the PSDa (NE) and PSDb (NW). The yellow cross is
      the reference point: the servo loop brings back the red point
      toward this position. Then, the metrology gondola is stabilized
      around a fixed position. The scale is in microns on the PSD. On
      these zooms each PSD has a field of view of $18\, \milli\meter$ at
      the metrology gondola level. Other technical
      windows used to adjust the PID ($K_p$, $K_I$, $K_d$) parameters
      or to create a matrix of interaction and the reference point are
      not displayed here. The ``position treuil'' (motorized winches
      positions) at the top of the window are used only for manual
      control.}
    \label{Windasserv}
  \end{figure}
\end{appendix}


\begin{appendix} 
\section{Discussion}
\label{discussion}
\subsection{Diluted Telescopes}

The science targets will
dictate the density of mirrors required in the pupil of the future interferometers because the
typical step of the diluted pattern imposes the ``Crowding Field of
View'' (FOV) limits (Lardi\`ere et al. \cite{Lardiere}, Koechlin
\cite{Koechlin}): if the
interferometer is very diluted (long distance between the
subapertures), images of large extended objects (angularly on the sky)
 reaches the crowding limit (Lardi\`ere et al. \cite{Lardiere}), and
is not observable. This limitation has already been reached on
the bigger stars with existing interferometers. For instance, some
Miras are not easily observable with long baseline interferometers
such as CHARA, Keck, or the VLTI.
For example, to obtain a direct image of a $30\,
\rm mas$ object with a $100\, \meter$ aperture interferometer,
at least 30 mirrors are required in the pupil.
Finally, to obtain a very high angular resolution (submilliarcsec), and many
``resels'' in the image, future interferometers (post-VLTI) will
require tens or hundreds of subapertures (rich uv coverage) over an
area of $\approx 100$--$300\,\meter$ in diameter, and a few telescopes distributed on
kilometer baselines (for very high angular resolution). For the same reasons, the ALMA submillimeter
 interferometer project consists of a giant array of
$12\,\meter$ antennas with baselines of up to $16 \,\kilo\meter$, and an
additional compact array of $7$--$12\,\meter$ antennas to enhance its
ability for imaging extended sources (Tarenghi \cite{Tarenghi}).

\noindent The Post-VLTIs will then be complex systems that will work
with many mirrors, actuators, and servo loops; they will be equipped
with adaptive optics and cophasing devices; their imaging properties
will resemble those of ELT telescopes but they will provide much higher angular
resolution. Their wavefront sensors will probably be similar to the
fringe tracking sensors adapted to the diluted pupils (Borkowski et
al. \cite{Borkowski}; Tarmoul et al. \cite{Tarmoul}). To build these
diluted telescopes made of hundreds of subapertures, the combined
knowledge of the scientific communities working on ELTs and
interferometers will be required.

\noindent We think that Carlina is a part of this new family of
instruments dedicated to unravel at high angular resolution the Universe: The diluted telescopes.
The MMT (Carleton \cite{Carleton}) and the large binocular telescope
interferometer (LBT) with its two $8.4\,\meter$ primaries, and its
$23\,\meter$ optical baseline could be precursors of these Diluted
Telescopes (Hinz \cite{Hinz}; Kim \cite{Kim}).

\noindent In order to achieve high imaging capabilities, and to be
more sensitive than present interferometers, a diluted telescope will
require the following characteristics:
\begin{itemize}
\item A relatively dense array of mirrors (rich uv coverage)
\item The optical train will be simplified to minimize the reflections
  between the primary mirrors and the focal instrument (probably
  without delay lines).
\item Focal instruments will be optimized to be sensitive and adapted
  to the science goals: AO adapted to the diluted telescopes, and
  coronagraphy will be implemented. The way to combine the telescopes
  will have to be optimized (see discussions in Patru et
  al. \cite{Patru}, Menut et al. \cite {Menut}).
\end{itemize}

\noindent From this point of view, Carlina, which works without
delay lines, is an ideal solution to recombine many mirrors. As a
next step, we shall propose to build a $100 \,\meter$ aperture diluted
telescope that we will call a LDT or very large
diluted telescope (VLDT) respectively for an equivalent surface
of a few meters square or $~50\, m^2$ (VLT surface).

\noindent A (V)LDT could be recombined in the densified mode
(Tallon \cite{Tallon}, Labeyrie \cite{Labeyriehyper}), or using any other technique
depending on the science goals (more studies are required).

\noindent It is also possible that
the post-ELTs will be partially diluted during their early construction phase.
It will then be also feasible to recombine the future ELTs,
with auxiliary telescopes on kilometric baselines, using single-mode fibers (Perrin et al. \cite{Perrin1})

\subsection{The site}\label{site}

The constraints for a (V)LDT's site are the same as for any modern
telescope: it should be located relatively high in the mountains (dry
air for IR observations), in a good weather area (weak cloud cover)
and with a favorable atmospheric turbulence ($seeing<1\arcsecond$,
slow turbulence). The ``Seeing'' of the valleys will have to be studied in
detail as it was done for example at Paranal (Dali et
al. \cite{Dali}). Nevertheless, the selection relies mainly on
topographic considerations. A valley oriented in
the E-W direction is required, with a nearly hemicylindrical
shape at the bottom (Fig. \ref{ligneniveau}).

\noindent The price tag of the project will mainly depend on the site
chosen to build a (V)LDT.  For this reason, we distinguish between two
kinds of site:
\begin{itemize}
\item The extremely good areas (ex: Paranal site) where we could
  propose only in a second stage to build a VLDT
  ($100$--$300\,\meter$ aperture) perhaps surrounded by telescopes
  over kilometric baselines.
\item The ``intermediate'' sites: the weather and turbulence are
  acceptable but generally not ideal; it is relatively easy to access
  the sites by road and to connect them to power, internet, etc.  The
  focal instrument will be optimized for such a turbulence.
\end{itemize}

\noindent 
We consider that the construction of a scientific demonstrator of LDT
with a $\approx 100\,\meter$  aperture is worth considering within the next ten years.
More studies are required to determine whether some intermediate sites
are really adapted to build an LDT for a reasonable cost. For a focal gondola at $f/2$ and an effective aperture of $100 \,\meter$, valleys of
  $400$ -- $600 \,\meter$ of curvature radius are required. By using Google Earth, and the ``Institut G\'eographique National'' (IGN) maps, we found in the southern Alps, nine intermediate sites that could have an adapted shape for a future LDT. Their preliminary characteristics are given in Table \ref{tablsite}.
  
  \begin{figure}[!h]
    \centering
    \includegraphics[width=8.8cm]{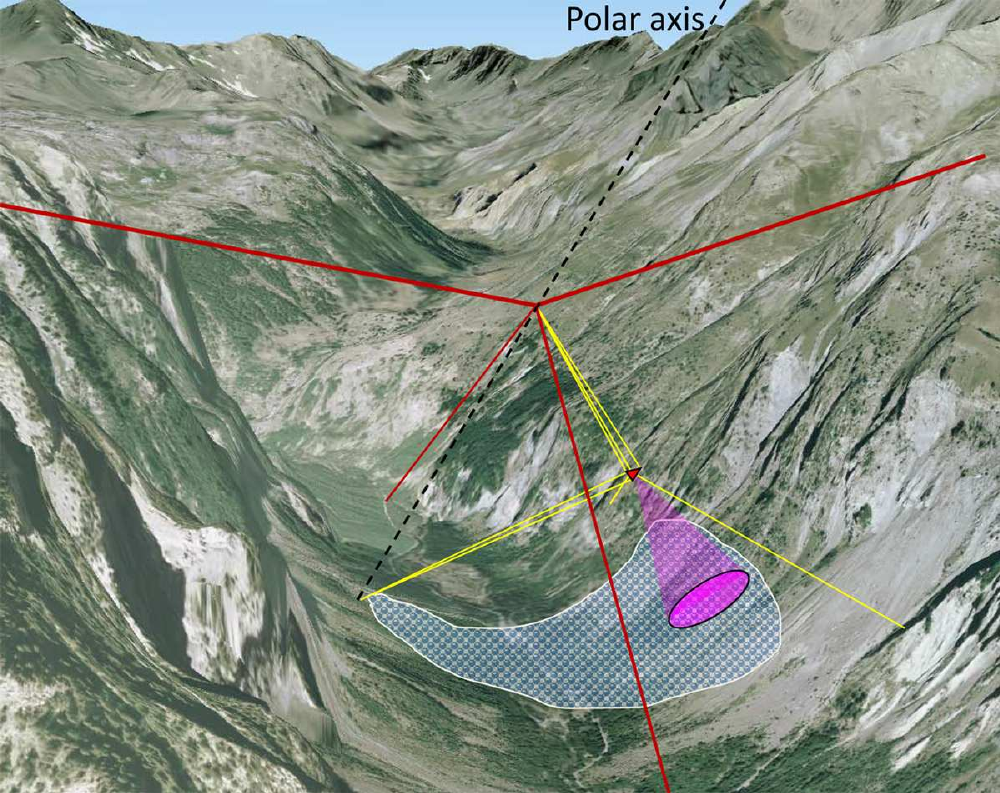}
    \caption{Drawing of a diluted telescope in the Valley. The pink circle represents the effective aperture of $120\,\meter$ at $F/2$.}
    \label{ligneniveau}
  \end{figure}

 \noindent Detailed studies are necessary to determine
precisely the properties (geometry, atmospheric seeing, snow and rock
avalanches, etc.) of potential sites. A study using a total station or/and aerial stereoscopic
  pictures will have to be done on the best sites. This study should determine the height of the primary mirror mounts in order to have a relatively homogeneous coverage of mirrors in the valley. 
 A study of the stability of the mounts in function of their heights, will be required.
  
  \noindent For example, a drawing of the installation of a diluted telescope in
  Freissini\`eres valley is presented in Fig.~\ref{ligneniveau}.  In such a
  site, it should be possible to track any star that passes above $40\degree$ from the horizon at meridian during about two hours.
  Geometrically, Freissini\`eres is a good site but it is surrounded
  by waterfalls, screes, etc.\\ 

\noindent Another possibility that should be explored is to put the
primary segments at the top of pylons (Fig.~\ref{dilutedPoteaux}) in
order to build a (V)LDT in a site with extremely good conditions of
atmospheric turbulence without topographic constraints. In this case,
the $10$--$200\,\meter$ pylons could be stabilized by a device equivalent to the system used for the
metrology gondola described in this paper.  The thermal expansion of a
$100\, \meter$ steel pylon would be around $1.2 \,\centi\meter$ for
$10\,\celsius$ variation.  We can then expect a drift of about $\approx
1\micro\meter/\second$ during the night.  A second servo loop stage, at
the top of each pylon would therefore be necessary to control the
piston and tip-tilt of each mirror.  Even though this might a priori
appear to complicate the design, one should keep in mind the
comparison with systems including tens of delay lines: it is probably
easier to stabilize a mirror around a fixed position than to control a
delay line moving at several millimeters per second. Moreover, the
internal metrology at the curvature center allows the coherencing of
the primary mirrors without any stellar source. 

\noindent More studies are also required installing a pylon array
with many cables (that must not touch each other even in strong wind
conditions). The impact on the local turbulence of such a massive
structure that gets cold during the night has also to be taken into
account. Digging an artificial crater in a flat terrain seems to be
impractical.\\


  \begin{table*}[!t]
  \center
\begin{tabular}{|p{2cm}|p{1.5cm}|p{1.5cm}|p{2.5cm}|p{1.5cm}|p{1.5cm}|}
\hline
\textbf{Sites Name} &  \textbf{Coordinates} & \textbf{Valley \newline bottom \newline altitude} & \textbf{Valley  \newline top altitude \newline N-S sides} & \textbf{Curvature radius}\newline ($\pm\,10\,m$) & \textbf{Baseline \newline at F/2}  \\
\hline
 La Mouti\`ere & $44^{\circ}\,19^{'}$ N \newline $6^{\circ}\,46^{'}$ E & 2070m & 2700m at north \newline 2600m at south & - & - \\
\hline
 Estrop & $44^{\circ}\,16^{'}$ N \newline $6^{\circ}\,31^{'}$ E & 2040m & 2720m \newline 2460m & 600m & 150m \\
\hline
Meollion & $44^{\circ}\,43^{'}$ N \newline $6^{\circ}\,17^{'}$ E & 1670m & 2800m \newline 2380m & - & - \\
\hline
Chaumeille & $44^{\circ}\,45^{'}$ N \newline $6^{\circ}\,18^{'}$ E & 1580m & 2670m \newline 2830m & - & - \\
\hline
Le Casset & $44^{\circ}\,49^{'}$ N \newline $6^{\circ}\,13^{'}$ E & 1150m & 2960m \newline 2300m & - & - \\
\hline
Freissini\`eres & $44^{\circ}\,44^{'}$ N \newline $6^{\circ}\,28^{'}$ E & 1370m & 3000m \newline 2450m & 450m & 112.5m \\
\hline
Deslioures & $44^{\circ}\,47^{'}$ N \newline $6^{\circ}\,27^{'}$ E & 1560m & 2700m \newline 3000m & 400m & 100m \\
\hline
Le Villard & $44^{\circ}\,50^{'}$ N \newline $6^{\circ}\,26^{'}$ E & 1270m & 2570m \newline 2450m & 600m & 150m \\
\hline
Les Etages & $44^{\circ}\,56^{'}$ N \newline $6^{\circ}\,16^{'}$ E & 1640m & 3200m \newline 2870m & & - \\
\hline
\end{tabular}
\caption{The main characteristics of sites found in the southern Alps. A minus sign is used in the boxes where we could not determine a value due to a lack of  numerical ground data. But, typically, all these valleys have about the same orientations (E/W), size, and shape.}
 \label{tablsite}
\end{table*}
  
\begin{figure*}[!t]
  \centering
  \includegraphics[width=15cm]{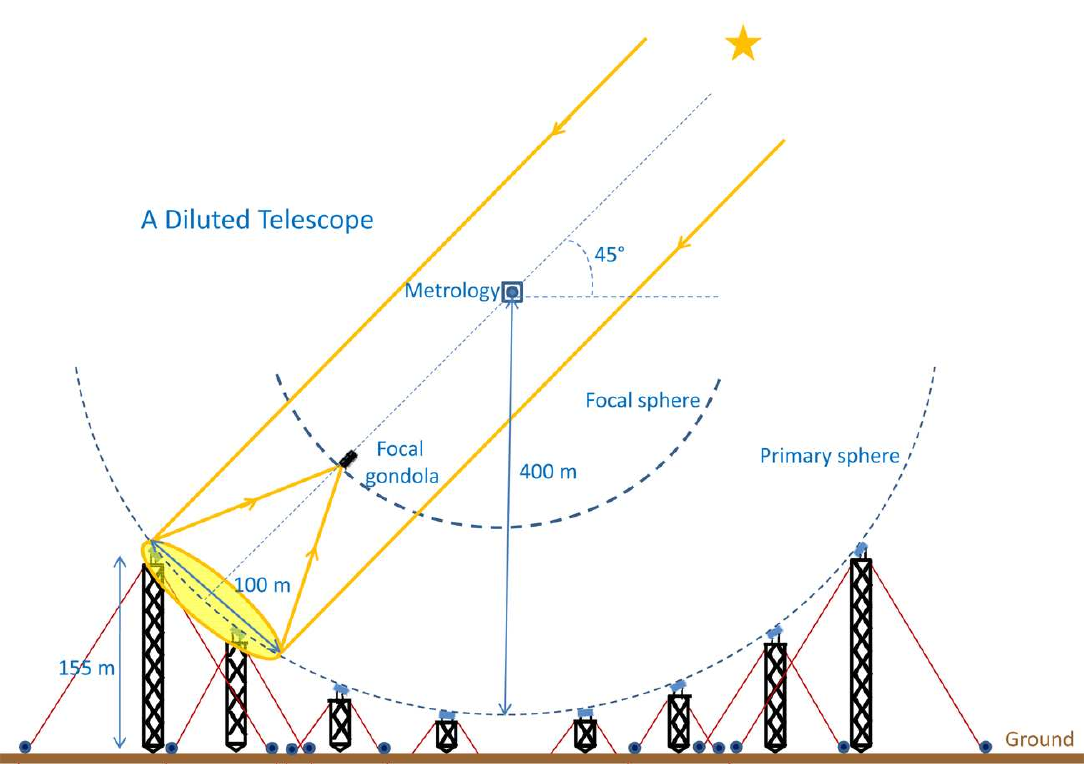}
  \caption{Schematic view of a Diluted Telescope using pylons to
    support the primary mirrors. The red cables are motorized to stabilize the top of the pylons using the same kind of servo
    loop system as that described in Sect. \ref{servoloop}.  A second mechanical
    stage stabilizes the primary mirrors within an accuracy of 1
    micron using a metrology at the curvature center as described in
    this article.}
  \label{dilutedPoteaux}
\end{figure*}

\end{appendix}

\begin{thebibliography}{}

\bibitem[2008]{Blazit} Blazit, A., Rondeau, X., Thi{\'e}baut, \'E.,
  Abe, L., Bernengo, J.-C., Chevassut, J.-L., Clausse, J.-M., Dubois,
  J.-P., Foy, R., Mourard, D., Patru, F., Spang, A., Tallon-Bosc, I.,
  Tallon, M., Tourneur, Y., Vakili, F. 2008, ApOpt., 47, 1141
\bibitem[2005]{Borkowski} Borkowski, V., Labeyrie, A., Martinache, F.,
  Peterson, D. 2005, A\&A 429, 747-753
  \bibitem[1979]{Carleton} Carleton, N. P. 1979, The MMT and the Future
  of Ground-Based Astronomy, Proceedings of a Symposium held to mark
  the dedication of the Multiple Mirror Telescope at the Mount Hopkins
  Observatory, Arizona on May 9, 1979. Edited by Trevor C. Weekes. SAO
  Special Report, 385, 37	
\bibitem[2009]{Courde} Courde, C., Lintz, M. and Brillet, A. 2009,
  Meas. Sci. Technol., 20, 127002
\bibitem[2010]{Courde2} Courde, C., Lintz, M. and Brillet, A. 2010
  Instrumentation Mesures Metrologie, 10(3-4), 81
  \bibitem[2010]{Dali} Dali Ali, W., Ziad, A., Berdja, A., Maire, J.,
  Borgnino, J., Sarazin, M., Lombardi, G., Navarrete, J., Vazquez
  Ramio, H., Reyes, M., Delgado, J. M., Fuensalida, J. J., Tokovinin,
  A., Bustos, E. 2010, A\&A, 524, A73
\bibitem[2011]{DeBecker} De Becker, M., Le Coroller, H. \& Dejonghe,
  J. 2011, in Proceedings of the 39th Li\`ege Astrophysical
  Colloquium, eds. G. Rauw, M. De Becker, Y. Naze, J.-M. Vreux \&
  P.M. Williams, BSRSL, 80, 486
  \bibitem[2010]{Dodorico} D'Odorico, S., Ramsay, S., Hubin, N., et
  al. 2010, The Messenger, 140, 17
\bibitem[2001]{Hinz} Hinz, P. 2001, American Astronomical Society,
  198th AAS Meeting, Bulletin of the American Astronomical Society,
  33, 859
  \bibitem[1997]{Jacubowiez} Jacubowiez, L., Sauer, H., Bernard, F.,
  Plantegenest, M.T., Avignon,T. 1997, Une exp\'erimentation
  p\'edagogique sur un syst\`eme d'optique adaptative. û
  SupOptique/IOTA û Colloque CETSIS, Orsay, 20-21 nov.1997
\bibitem[2010]{Kim} Kim, J., Hinz, P., Durney, O., Connors, T.,
  Montoya, M., Schwab, C. 2010, Proceedings SPIE, Testing and
  alignment of the LBTI, Optical and Infrared Interferometry II.,
  V. 7734, 77341W-77341W
\bibitem[2003]{Koechlin} Koechlin L., 2003, EAS, 8, 349
\bibitem[1996]{Labeyriehyper} Labeyrie, A. 1996, A\&As, 118, 517
\bibitem[2003]{Lardiere1} Lardi\`ere, O., Labeyrie, A., Mourard, D.,
  Riaud, P., Arnold, L., Dejonghe, J., Gillet, S. 2003, Interferometry
  for Optical Astronomy II. Edited by Wesley A. Traub . Proceedings of
  the SPIE, 4838, 1018-1027
\bibitem[2007]{Lardiere} Lardi\`ere, O., Martinache, F., Patru, F. 2007
  MNRAS, 375, 977
\bibitem[2004]{Lecoroller} Le Coroller, H., et al. 2004, A\&A, 426,
  721 (Paper I)
\bibitem[2010]{lecoroller2} Le Coroller, H., Dejonghe, J., Regal, X.,
  De Becker, M., Sottile, R., Ricci, D., Meunier, J.P., Guillaume, C.,
  Carole, C., Blazit, A., Clausse, J.M., Labeyrie, A., Mourard, D., Le
  VanSuu, A., Boer, M. 2010, The European Week of Astronomy and Space
  Science, in proceedings of JENAM 2010, Lisbon 6-10 September 2010,
  eds. A. Moitinho, E. Amores, V. Arsenijevic, J. Ascenso, R. Azevedo,
  P. 95
\bibitem[2009]{Meisenheimer} Meisenheimer, K. 2009, Science with the
  VLT in the ELT Era, Astrophysics and Space Science Proceedings,
  Volume . ISBN 978-1-4020-9189-6. Springer Netherlands, 507
\bibitem[2008]{Menut} Menut, J.L., Valat, B., Lopez, B., Schmider,
  F.X., Vakili, F. et al. 2008, ApJ, 686, 1514
  \bibitem[1996]{Mertz} Mertz, L. 1996, Excursions in Astronomical Optics (New York:
Springer-Verlag, Inc.)
    \bibitem[2009]{Mourard} Mourard, D., Clausse, J.M., Marcotto, A. et
  al. 2009, A\&A, 508, 1073
\bibitem[1995]{Orndoff} Orndoff, E. 1995, NASA, Lyndon B. Johnson Space Center
Houston Texas, "Development and Evaluation of Polybenzoxazole Fibrous Structures", Technical Memorandum 104814
\bibitem[2008]{Patru} Patru, F., Mourard, D., Clausse, J.M., et
  al. 2008, A\&A, 477, 345
\bibitem[2009]{Patru2} Patru, F., Tarmoul, N., Mourard, D. and
  Lardi{\`e}re, O. 2009, MNRAS, 395, 2363
  \bibitem[2004]{Perrin} Perrin, G., Lai, O., Woillez, J., et al. 2004,
  American Astronomical Society Meeting 204, Bulletin of the American
  Astronomical Society, 36, 982
\bibitem[2000]{Perrin1} Perrin, G. et al. 2000, SPIE, ``A fibered
  large interferometer on top of Mauna Kea: OHANA, the Optical
  Hawaiian Array for Nano-radian Astronomy'', 4006, 708
\bibitem[2007]{Petrov} Petrov, R.G., Malbet, F., Weigelt, G. et
  al. 2007, A\&A, 464, 1
\bibitem[2010]{Ridgway} Ridgway, S., Glindemann, A. 2010, EAS
  Publications Series, 40, 235-243
\bibitem[1992]{Tallon} Tallon, M. and Tallon-Bosc, I. 1992 A\&A, 253, 641
\bibitem[2008]{Tarenghi} Tarenghi, M. 2008, Astrophys Space Sci, 313,
  1-7
\bibitem[2010]{Tarmoul} Tarmoul, N., Mourard, D., H{\'e}nault, F.,
  Clausse, J.M., Girard, P., Marcotto, A., Mauclert, N., Spang, A.,
  Rabbia, Y., Roussel, A. 2010, SPIE, Presented at the Society of
  Photo-Optical Instrumentation Engineers (SPIE) Conference, 7734, 68
\bibitem[2000]{Weigelt} Weigelt, G., Mourard, D., Abe, L., Beckmann,
  U., Chesneau, O., Hillemanns, C., Hofmann, K., Ragland, S., Schertl,
  D., Scholz, M., Stee, P., Thureau, N., Vakili, F. 2000,
  Interferometry in Optical Astronomy, Pierre J. Lena; Andreas
  Quirrenbach; Eds., Proc. SPIE, 4006, 617
\bibitem[2005]{Woillez} Woillez, J., Perrin, G., Lai, O. 2005, AAS,
  Bulletin of the American Astronomical Society, 37, 1308
\end{thebibliography}
\end{document}